\documentclass[fleqn,usenatbib]{mnras}

\usepackage{newtxtext,newtxmath}

\usepackage[T1]{fontenc}

\DeclareRobustCommand{\VAN}[3]{#2}
\let\VANthebibliography\thebibliography
\def\thebibliography{\DeclareRobustCommand{\VAN}[3]{##3}\VANthebibliography}

\usepackage{graphicx}	
\usepackage{gensymb}
\usepackage{amsmath}	
\usepackage{multirow}
\newcommand{\ascii}[1]{{\fontfamily{qcr}\selectfont #1}}

\title[X-ray selected rQSO]{SDSS-V: Revealing a weak accretion state in X-ray selected red quasars.}

\author[P. Guetzoyan et al.]{
Paloma Guetzoyan,$^{1}$\thanks{E-mail: paloma.guetzoyan@ed.ac.uk}
James Aird,$^{1}$
Amy L. Rankine,$^{1}$
Stephanie M. LaMassa,$^{2}$
Peter Breiding,$^{3}$
Mara Salvato,$^{4}$
\newauthor
Johannes Buchner,$^{4}$
Zsofi Igo,$^{4,5}$
Roberto J. Assef,$^{6}$
Hector Ibarra-Medel,$^{7}$
Catarina Aydar,$^{4}$
\newauthor
Castalia Alenka Negrete,$^{7}$
Claudio Ricci,$^{8,9}$
W. N. Brandt,$^{10,11,12}$
Dong-Woo Kim,$^{13}$
Dominika Wylezalek,$^{14}$
\newauthor
Scott F. Anderson,$^{15}$
Donald P. Schneider,$^{16,17}$
Delvin Demke,$^{14}$
and Anton M. Koekemoer$^{2}$
\\
$^{1}$Institute for Astronomy, University of Edinburgh, Royal Observatory, Edinburgh EH9 3HJ, UK\\
$^{2}$Space Telescope Science Institute, 3700 San Martin Drive, Baltimore, MD 21218, USA\\
$^{3}$Department of Physics, Applied Physics, and Astronomy,
Binghamton University, Binghamton, NY 13902, USA\\
$^{4}$Max-Planck-Institut f\"ur extraterrestrische Physik (MPE), 
Gie{\ss}enbachstra{\ss}e 1, 85748 Garching bei M\"unchen, Germany\\
$^{5}$Exzellenzcluster ORIGINS, Boltzmannstr. 2, 85748 Garching, Germany\\
$^{6}$Instituto de Estudios 
Astrof\'isicos, Facultad de Ingenier\'ia y Ciencias, Universidad Diego Portales, Av. Ej\'ercito Libertador 441, Santiago, Chile\\
$^{7}$Universidad Nacional Autónoma de México, Instituto de Astronomía, AP 70-264, CDMX 04510, México\\
$^{8}$Department of Astronomy, University of Geneva, ch. d'Ecogia 16, 1290, Versoix, Switzerland\\
$^{9}$Instituto de Estudios Astrof\'isicos, Facultad de Ingenier\'ia y Ciencias, Universidad Diego Portales, Av. Ej\'ercito Libertador 441, Santiago, Chile\\
$^{10}$Department of Astronomy and Astrophysics, 525 Davey Lab, The Pennsylvania State University, University Park, PA 16802, USA\\
$^{11}$Institute for Gravitation and the Cosmos, The Pennsylvania State University, University Park, PA 16802, USA\\
$^{12}$Department of Physics, 104 Davey Laboratory, The Pennsylvania State University, University Park, PA 16802, USA\\
$^{13}$Center for Astrophysics | Harvard \& Smithsonian, 60 Garden Street, Cambridge, MA 02138, USA\\
$^{14}$Astronomisches Rechen-Institut, Zentrum für Astronomie der Universit\"at Heidelberg, M\"unchhofstr. 12-14, 69120 Heidelberg, Germany\\
$^{15}$Department of Astronomy, University of Washington, Box 351580, Seattle, WA 98195, USA\\
$^{16}$Department of Astronomy and Astrophysics, The Pennsylvania State University, University Park, PA 16802, USA\\
$^{17}$Institute for Gravitation and the Cosmos, The Pennsylvania State University, University Park, PA 16802, USA
}

\date{Accepted XXX. Received YYY; in original form ZZZ}

\pubyear{\the\year{}}

\begin{document}
\label{firstpage}
\pagerange{\pageref{firstpage}--\pageref{lastpage}}
\maketitle

\begin{abstract}
Red quasars (rQSOs) have been recently recognized as a short-lived, early stage in the evolutionary cycle of Active Galactic Nuclei (AGN), with fundamental differences in their intrinsic properties compared to typical blue quasars (bQSOs). In this work, we present the first large X-ray sample of 380 rQSOs, selected from the \textit{eROSITA}/SDSS-V collaboration, providing uniform X-ray detection with optical spectroscopy across half the sky, in the German hemisphere of \textit{eROSITA}. We combine X-ray imaging, optical spectroscopy, and multi-wavelength photometry to fully probe the accretion, absorption, spectroscopic, and host properties of rQSOs and compare them to a bQSO sample. Independent Component Analysis methods are used to reconstruct optical spectra in a data-driven and non-parametric approach, while accounting for dust reddening and host contamination. rQSOs are intrinsically X-ray weak compared to bQSOs, with a higher fraction found at low X-ray luminosities (over 50$\%$ of the rQSO sample have $L_X < 10^{43.5}\, \rm erg \, s^{-1}$). We investigate the relative X-ray strength of rQSOs by measuring the spectral slope indicator $\alpha_{OX}$. Despite their suppressed X-ray emission, rQSOs are not optically faint, but show low $\alpha_{OX}$ values and higher average dust-corrected optical luminosities, indicating weak X-ray emission relative to their bright optical continua. X-ray spectral measurements reveal large gas column densities relative to optical reddening due to dust, implying that X-ray absorption could arise from dust-free gas close to the supermassive Black Hole (BH) rather than a classical dusty torus, while the dust responsible for optical reddening likely resides on larger host-galaxy scales or is associated with dusty gas carried in disc winds. rQSOs trace a phase of suppressed BH assembly relative to stellar mass growth, suggesting that they represent a distinct evolutionary stage where BH accretion is suppressed while the host galaxy continues to grow.

\end{abstract}




\section{Introduction}

Supermassive Black Holes (BHs) of mass $M_{\rm BH}= 10^{5}-10^{10}\, \rm M_{\odot}$ are now recognized as central engines in the evolution of their host galaxies. Most, if not all, massive galaxies harbour a BH at their centre \citep[e.g.][]{magorrian_demography_1998, kormendy_supermassive_2001}. Strong correlations, such as relations between BH mass, galaxy bulges, total stellar mass or the stellar velocity dispersion \citep[$M_{\rm BH} - M_{\rm bulge}$, $M_{\rm BH}-M_{\rm star}$ and $M_{\rm BH}-\sigma$ relations, e.g., ][]{kormendy_inward_1995,
gebhardt_black_2000,ferrarese_fundamental_2000, tremaine_slope_2002, kormendy_coevolution_2013,heckman_coevolution_2014, 
reines_relations_2015, shankar_selection_2016, greene_intermediate-mass_2020}, indicate the existence of co-evolution mechanisms between BH growth and the mass assembly of their galaxies, bridging their connection across scales. Although several scenarios could lead to finding such correlations \citep[common gas supplies, merger averaging, e.g.,][]{breiding_powerful_2024}, this connection is thought to be driven by the remarkable amount of energy released in the vicinity of BHs, when they enter the Active Galactic Nuclei (AGN) phase \citep{shapiro_black_1983}, powered by accretion of gas onto their central BH. AGN can produce strong radiation across the full electromagnetic spectrum \citep[e.g.][for a recent review]{padovani_active_2017} including X-ray emission in the corona and UV/optical emission from the accretion disc, while also driving outflows of material observed at radio wavelengths. The brightest AGN, known as quasars, can become so luminous that they completely outshine their host galaxies, appearing as the most luminous point-like sources in the Universe. Gas accretion during the AGN phases is thought to be the dominant channel for BH growth, yet, the physical mechanisms regulating BH accretion remain long-standing questions. Since the role and efficiency of AGN feedback are intrinsically linked to BH accretion \citep{hopkins_unified_2006,fabian_observational_2012}, we cannot fully understand the co-evolution of galaxies and BHs without first determining how and when AGN are triggered, and under which conditions BHs grow most efficiently.

We now know that AGN are not static, stable objects, but instead cycle through obscured and unobscured stages as accretion rises and fades \citep{hickox_black_2014}. Gas inflows can efficiently fuel the BH but also simultaneously obscure the nucleus on various scales \citep{ricci_close_2017,ricci_bass_2022}. With stronger accretion, AGN-driven winds and radiation pressure start expelling the surrounding dust, transitioning from a dust-enshrouded quasar to a more unobscured blue quasar phase, where the accretion disc is revealed. A widely adopted explanation for obscuration is given by an orientation-based unified model \citep{antonucci_unified_1993}. In this model, the diversity of AGN properties arises mainly from the observer's viewing angle to the optically-thick dusty torus surrounding the accretion disc and not from capturing the AGN at different evolutionary stages. Within this orientation picture, obscured and unobscured AGN share the same underlying intrinsic properties, and only differ due to geometric effects.

Red quasars \citep[rQSOs, e.g.,][]{webster_evidence_1995,richards_red_2003,glikman_first2mass_2004,glikman_first-2mass_2012,banerji_heavily_2012,banerji_heavily_2015} have emerged as a key population, showing a broad-line optical spectrum with heavily reddened continuum. Due to the detection of broad lines, their red colours cannot be explained by orientation effects only and early studies already questioned whether their reddening arises purely from obscuration due to a dusty torus or from dust on larger scales \citep[e.g.,][]{urrutia_evidence_2008}. In an orientation scenario, differences between obscured and unobscured AGN would arise purely from our line-of-sight intersecting the dusty torus or not, predicting that rQSOs are type-1 AGN, as broad emission lines are found in their optical spectra, but viewed through mild obscuration from dust along the torus edge. If orientation were the main effect, rQSOs should have the same intrinsic outflow and accretion properties as blue quasars (bQSOs). However, recent studies, challenge this orientation-based picture of rQSOs, demonstrating that systematic differences persist between rQSOs and typical type-1 bQSOs even after correcting for orientation effects, thus favouring an evolutionary scenario \citep{glikman_first-2mass_2012,banerji_heavily_2015,klindt_fundamental_2019,fawcett_fundamental_2020}. 

In such a scenario, rQSOs represent a transitional phase in the co-evolution of BHs and their hosts \citep[see][for a recent review]{alexander_what_2025}. This scenario begins with a heavily dust-enshrouded type-2 AGN phase, where both the BH and host galaxy are rapidly growing, fuelled by large gas inflows triggered by mergers or interactions. As accretion onto the BH intensifies, AGN-driven feedback in the form of powerful winds and outflows begins to clear the surrounding dust and gas in a short-lived “blow-out” phase -- the rQSO phase. Once this obscuring material is dispersed, the nucleus is revealed as a typical unobscured blue type-1 AGN (or bQSOs). In this scenario, dust and gas play a central role: either gas-fuelled accretion or suppressing and obscuring BH growth through feedback effects. Nuclear dust not only regulates BH growth but also shapes the observed properties of AGN, which makes it key information on the geometry of the obscuring material and the intrinsic accretion physics. Recent radio studies provide notable support for rQSOs being a transitional phase. 
\citealt{klindt_fundamental_2019} has shown that rQSOs have a higher radio detection fraction compared to a control sample of bQSOs, suggesting enhanced radio activity in this population. However, despite this finding, \citealt{fawcett_fundamental_2022,fawcett_striking_2023} reveal that rQSOs tend to still be in the radio-quiet regime with compact radio morphologies, suggesting the enhanced radio detection fractions is due to small scale winds from the accretion disk or young compact radio jets interacting with the interstellar medium (ISM) of the galaxy. This out-flowing material is thought to contain the dust that is responsible for the reddening of optical spectra and colours of rQSOs.

However, the physical origin of this dust, as well as whether there are intrinsic differences between these sources and the wider quasar population, remain unclear, highlighting the need for a systematic and homogeneous study of these sources. In particular, it is still unknown from radio-selected samples, if dust properties are directly linked to altered accretion states in rQSOs. X-ray studies provide a powerful probe to answer these questions. The hard X-ray emission traces the hot inner corona, providing insights into the accretion state of the BH, while absorption of soft X-rays provides a tracer of the overall line-of-sight density of gas and can thus provide complementary probe of the location of the dusty gas responsible for the reddening of the optical emission in rQSOs \citep[e.g.,][]{haardt_two-phase_1991,done_intrinsic_2012,brandt_cosmic_2015}.

While most rQSO studies have been focused on their nuclear and feedback properties, some works aimed at examining rQSO hosts directly  to constrain morphologies, merger-fractions or host-scale dust structures \citep[e.g.,][]{glikman_major_2015,zakamska_host_2019}. However, the connection to the host galaxy has often been limited by poor imaging, wavelength coverage and differences in timescales on which changes operate, making it difficult to separate AGN and host dust emission. Within an evolutionary scenario, large gas reservoirs associated with massive star-forming galaxies can fuel rapid BH growth while providing substantial dust content on host-scale, potentially obscuring the optical emission of the nucleus. Thus, dust reddening could also arise from galaxy-scale dust structures within the ISM of the host, or dust carried in large-scale winds \citep{calistro_rivera_multiwavelength_2021}. Disentangling nuclear and host contributions is thus essential to understand whether rQSOs suffer from nuclear or large-scale absorption, as well as robustly constrain stellar masses and star formation rates. 

While previous studies suggest that rQSOs are key to understanding obscuration, feedback processes, and possibly new accretion states, well-defined and homogeneous samples of rQSOs remain sparse, particularly the X-ray-selected samples. Current X-ray studies all suffer from several limitations: X-ray follow-up observations of optically-selected rQSOs, often from small, heterogeneous samples, and inconsistent definitions of rQSOs that prevent comparisons across samples and redshift. \citealt{lamassa_peering_2016} and \citealt{glikman_peering_2017} presented a combined sample of four FIRST-2MASS \citep{skrutskie_two_2006} rQSOs with targeted X-ray spectroscopy and found all sources to be either X-ray weak or over-luminous in the infrared. However, small samples and the limitations cited above have prevented robust conclusions and, in some cases, led to contradictory results regarding the intrinsic nature of rQSOs. For instance, \citealt{goulding_high-redshift_2018} study 11 rQSOs with X-ray data from the \textit{XMM} and \textit{Chandra} telescopes, but only two have sufficient X-ray counts for detailed analysis. They find evidence of substantial obscuring material along the line-of-sight, measuring line-of-sight absorption columns of $\log_{10} \, N_{\rm H} \sim 23-24 \rm \, cm^{-2}$, but after absorption correction, the intrinsic X-ray luminosities were consistent with normal bQSOs with matched optical luminosity, suggesting no intrinsic X-ray weakness. In contrast, \citealt{ma_evidence_2024} selected 40 rQSOs from an optical sample, to target with X-ray follow-up observations with \textit{Chandra}. Most of their targets were undetected, while those with X-ray detections systematically exhibit weaker X-ray emission than expected, compared to bQSOs.

Thus, a homogeneous and statistically significant X-ray selected sample is required to accurately measure intrinsic accretion properties of rQSOs and improve the evolutionary picture of AGN. X-ray observations are capable of penetrating the dust that obscures quasars at optical and IR wavelengths, and is the selection method the least biased towards obscuration \citep{hickox_obscured_2018}. X-rays provide an unbiased probe of accretion activity, allowing the detection of rQSOs that would otherwise be missed or biased toward only the brightest sources in other surveys. The emergence of recent all-sky X-ray surveys with \textit{eROSITA} \citep{merloni_erosita_2012,merloni_srgerosita_2024} combined with the SDSS-V Black Hole Mapper program \citep{kollmeier_sloan_2026} provides for the first time uniform X-ray detection of quasars across half the sky with high-resolution optical spectroscopic observations. This dataset enables a systematic study of X-ray selected rQSOs to connect their X-ray emission, dust reddening, and host properties. 

In this paper, we use a large sample of X-ray selected quasars from \textit{eROSITA} with optical spectroscopy with SDSS-V to investigate the differences in accretion, host, and spectroscopic properties between rQSOs and bQSOs. Section \ref{sec:data} describes the datasets: \textit{eROSITA} X-ray imaging, SDSS-V optical spectroscopy, and multi-wavelength photometry. Section \ref{sec:method} details our photometric selection of rQSOs, our novel non-parametric approach for spectroscopic reconstruction while accounting for host contribution and dust reddening, and our SED modelling method. We present our various results in Section \ref{sec:results}, investigating the differences between rQSOs and bQSOs in terms of X-ray emission, spectroscopic and host properties. We discuss our results in Section \ref{sec:discussion}, by comparing our work to the literature and exploring scenario for dust origin in rQSOs and its implication for an evolutionary scenario. We conclude and summarize our results in Section \ref{sec:conclusions}. 

Throughout this work, we assume standard cosmology parameters ($H_0 = 70$\,km\,s$^{-1}$\,Mpc$^{-1}$, $\Omega_m = 0.3$, $\Omega_\Lambda = 0.7$) and use photometry defined in the AB system.

\section{Data}
\label{sec:data}

In this work, we define a parent sample of X-ray selected quasars, within which we subsequently identify a subset of reddened QSO on the basis of observed $g-r$ colours (see Section \ref{sec:selection}). The parent sample comprises all sources spectroscopically classified as QSO\footnote{The QSO spectroscopic classification comes from the SDSS pipeline based on the presence of broad emission lines} within the eROSITA targets in the SDSS spectroscopic follow-up programme (based on eRASS1 detections). After removing duplicates with the lowest signal-to-noise ratio and sources flagged with a redshift warning, our sample contains 8,625 sources, i.e., X-ray AGN with robust optical spectroscopic and broad UV/MIR photometric coverage.

In this section, we describe the X-ray and optical datasets used in this paper, with the X-ray data coming from \textit{eROSITA}, the Black Hole Mapper (BHM) programme of the SDSS-V providing ongoing optical spectroscopy of the X-ray detected sources, and the multi-wavelength photometric catalogue we compile after matching optical positions. Section \ref{sec:erosita} describes the X-ray photometric data from the eROSITA all-sky survey and the deeper eROSITA performance verification field eFEDS (eROSITA Final
Equatorial Depth Survey field), Section \ref{sec:SDSS} details the optical spectroscopic follow-up of the X-ray sample with SDSS, and Section~\ref{sec:photometry} lists the various the photometric surveys from UV to MIR we cross-matched to our sample.

\subsection{X-ray imaging with eROSITA}
\label{sec:erosita}
The eROSITA/SRG mission \citep{merloni_erosita_2012, predehl_erosita_2021,sunyaev_srg_2021}, launched in 2019, has conducted the first uniform soft X-ray all-sky survey since ROSAT \citep{trumper_rosat_1982, boller_second_2016} achieving significantly improved sensitivity. In this work, we exploit eROSITA data from the western Galactic hemisphere (Galactic longitudes $180^{\circ} < l < 360^{\circ}$) provided by the German eROSITA collaboration (eROSITA-DE). All statements regarding survey performance refer to the data in this hemisphere. Over the six-month period required to complete the first all-sky survey, eROSITA detected roughly 900,000 point sources in the soft $0.2 - 2.3\, \rm keV$ band; $75\%$ of events expected to be associated with AGN emission. In the soft band, eROSITA reaches a flux limit of $5\times 10^{-14}\, \rm erg \, s^{-1}.cm^{-2}$ for a single all-sky scan and $7\times 10^{-13}\, \rm erg \, s^{-1}.cm^{-2}$ in the hard band $(2.3 - 5\, \rm keV)$ \citep[eRASS1]{merloni_srgerosita_2024}. These limits represent a $\sim 5 \times$ improvement over ROSAT's sensitivity, while extending into higher energy ranges, and has already doubled the known X-ray source population. 

To obtain a large, uniformly-selected sample for this work, we require a detection in the soft (0.2$-$2.3~keV) eROSITA band where the sensitivity is greatest. Here we restrict our analysis to the publicly available eRASS1 data release \citep[]{merloni_srgerosita_2024,salvato_counterpart_2025}, as well as eFEDS data \citep{brunner_erosita_2022,salvato_erosita_2022,aydar_erosita_2025}, the deeper calibration field of eROSITA of $140\, \rm deg^2$, located at 126$\degree<\alpha_{J2000}<$146$\degree$ and -3$\degree<\delta_{J2000}<$6$\degree$. Our final sample consists of X-ray sources from those eROSITA public releases that were selected for spectroscopic observations by SDSS. Since targeting is still ongoing, we limited our sample to the latest available SDSS data release DR19 \citep[][see Section \ref{sec:SDSS} for details of the eROSITA/SDSS catalogue used in this work]{sdss_collaboration_nineteenth_2025}.
Details of the point-source detection procedure, photometry, and subsequent flux estimates (including a correction for the impact of the Point Spread Function) are given by \citealt{predehl_erosita_2021} and \citealt{maitra_sn_2022}. We derive X-ray luminosities from those observed 0.2--2.3~keV fluxes, $f_{obs}$, corrected for Galactic absorption using $N_\mathrm{H,Gal}=3\times10^{20}$~cm$^{-2}$ for all sources as the typical value  (\citealt{merloni_srgerosita_2024}), using the following approach. Assuming that fluxes follow a powerlaw spectral density of photon index $\Gamma = 1.9$, which is the canonical assumption value assumed for AGN and is consistent with the $\Gamma \sim 2$ typically found in \textit{eROSITA} AGN, we convert $f_{obs}$ into rest-frame 0.2--2.3~keV fluxes $f_{RF} = f_{obs}\times (1+z)^{\Gamma - 2}$, where $z$ is the redshift inferred from the optical spectrum obtained by SDSS (see Section \ref{sec:SDSS} below). Rest-frame fluxes are converted into rest-frame luminosities in the soft band such that $L_{0.2-2.3\, \rm keV} = 4\pi D_L^2(z)f_{RF}$, with $D_L$ being the luminosity distance derived under the standard cosmology assumptions.


\subsection{SDSS-V optical spectroscopic follow-up}
\label{sec:SDSS}
The Sloan Digital Sky Survey \citep[SDSS][]{york_sloan_2000} has provided the broadest sky coverage of any ground-based optical spectroscopic survey for decades, and has recently expanded with SDSS-IV, using now two observing facilities, one in each hemisphere, Las Campanas Observatory in Chile \citep{bowen_optical_1973}, and the Apache Point Observatory in New Mexico \citep{gunn_25_2006}.
The SDSS BOSS spectrographs \citep{smee_multi-object_2013} have a spectral resolution of $R = 1560-2650$ and a wavelength coverage with good throughput from 3650\r{A} to 9500\r{A}. 

SDSS-V \citep{kollmeier_sloan_2026} and \textit{eROSITA} collaborate through the SPIDERS project (SPectroscopic IDentification of ERosita Sources, \citealt{dwelly_spiders_2017, comparat_final_2020}), where X-ray detected sources are targeted for the Black Hole Mapper programme (Anderson et al. 2026 in prep.), providing optical observations for a large sample of about 300,000 quasars detected by the X-ray probe, providing robust spectroscopic classification and redshift measurement of the X-ray sources to a limiting magnitude of about $i\sim21.5$. 
This ambitious spectroscopic survey targets eROSITA AGN and clusters across $10,000\, \rm deg^2$ for eRASS scans, as well as X-ray sources in eFEDS. All SPIDERS targets from the first eROSITA data release \citep{merloni_srgerosita_2024} / SDSS DR19 \citep{sdss_collaboration_nineteenth_2025} are listed in a Value Added Catalog\footnote{\url{https://www.sdss.org/dr19/data_access/value-added-catalogs/?vac_id=10004}} (VAC) released alongside DR19 (see Section 7.2 of \citealt{sdss_collaboration_nineteenth_2025}). This VAC contains both optical and basic X-ray data information, such as observed fluxes for a total of 8625 unique quasars from both eFEDS and eRASS1 with reliable spectroscopic redshifts in DR19. Some multi-wavelength photometric data from GAIA DR3 \citep{gaia_collaboration_gaia_2023}, unWISE \citep{meisner_unwise_2019}, GALEX \citep{bianchi_galaxy_2014}, and 2MASS \citep{mcmahon_first_2013} are also included in the VAC, but we supplement it to increase the wavelength coverage as described in the following sub-section.

We correct each spectrum for Galactic extinction. The amount of reddening depends on the line of sight dust column through our Galaxy. We derive the local extinction $E(B-V)$ at the coordinates of each source using the dust maps from \citealt{schlegel_maps_1998} implemented in the \ascii{dustmaps} \ascii{python} module \citep{m_green_dustmaps_2018} and remove the corresponding reddening from the observed spectrum using the extinction curve of \citealt{fitzpatrick_correcting_1999}.

\subsection{Optical/IR photometry}
\label{sec:photometry}
In this work, we use optical photometry, specifically the $g$ and $r$ magnitudes from Legacy Survey DR10 to select rQSO (more details in Section \ref{sec:selection}). Moreover, broad photometric coverage is essential for robust host-galaxy characterization through Spectral Energy Distribution (SED) fitting (see Section \ref{sec:SED}). To complement our optical spectroscopic data, we cross-match our sample with wide photometric surveys spanning UV to MIR wavelengths. The cross-matching is done based on optical positions within a $1''$ radius to be conservative in case of rounding errors between the two optical positions, but all matches are found to be within less than $0.1''$.

We retrieve UV photometry from the SDSS DR19/eRASS1 VAC (see Section \ref{sec:SDSS} for more detail), taken from the GALEX survey (data release 6/7\footnote{\url{https://galex.stsci.edu/GR6/}}), providing FUV and NUV magnitudes. For the remaining bands, we perform an independent cross-match between optical positions to the following surveys to increase photometric coverage. The optical bands $g,r,i,z$ are taken from the Legacy Survey data release 10\footnote{\url{https://www.legacysurvey.org/dr10/catalogs/}}, which compiles photometric data within the footprint of the Dark Energy Spectroscopic Instrument survey \citep[DESI]{desi_collaboration_overview_2022} from the Mayall z-band Legacy Survey MzLS \citep[]{silva_mayall_2016}, the Dark Energy Camera Legacy Survey DECaLS \citep[]{blum_decam_2016} and the Beijing-Arizona Sky Survey BASS \citep{zou_project_2017} programmes. We also include NIR photometry from the Two Micron All Sky Survey \citep[2MASS]{skrutskie_two_2006} with $J, H, K_s$ as well as the two wide field surveys from VISTA, VIKING \citep[]{edge_vista_2013}, and VHS \citep[]{mcmahon_first_2013}. MIR photometry is taken from the WISE all-sky survey \citep[]{wright_wide-field_2010} providing $W1,W2,W3,W4$ magnitudes. Our photometric data span a wavelength coverage of $\sim 150$ \,nm -- 22\,microns, ensuring reliable SED modelling across a broad range of rest-frame wavelengths.

\section{Methods}
\label{sec:method}

\begin{table}
    \centering
    \begin{tabular}{c|c|}
    \multicolumn{2}{|c|}{Number count}\\
    \hline
    & Parent sample \\
    \hline
    ZWARNING$=0$ & 23827 \\
    \hline
    SPIDERS targets & 21995  \\
    \hline
    Unique sources & 12138 \\
    \hline
    QSO classification & 8625  \\
    \hline
    Detection in $W1$ & 8237  \\
    \hline
    $S/N > 1.5$  & 4143 (3763 bQSOs and 380 rQSOs) \\
    \hline

    \end{tabular}
    \caption{Number count of sources for the DL1 parent sample after each cut.}
    \label{tab:ncount}
\end{table}

This section describes our method to identify the red QSO (rQSO) sample, based on a photometric selection criterion, from within our parent sample of X-ray--selected QSO identified by SDSS-V (Section \ref{sec:selection}). Section \ref{sec:specfit} presents our modelling of the SDSS optical spectra using a spectral reconstruction process that accounts for host-galaxy contributions, measure the intrinsic reddening of the quasar light, and estimate the intrinsic properties of the rQSO. Section \ref{sec:SED} describes our modelling of the broader photometric Spectral Energy Distribution (SED) used to estimate host stellar masses and provide further insights into the properties of rQSOs.

\subsection{Selecting the red quasar sample}
\label{sec:selection}
\begin{figure}
    \centering
    \includegraphics[scale=0.5]{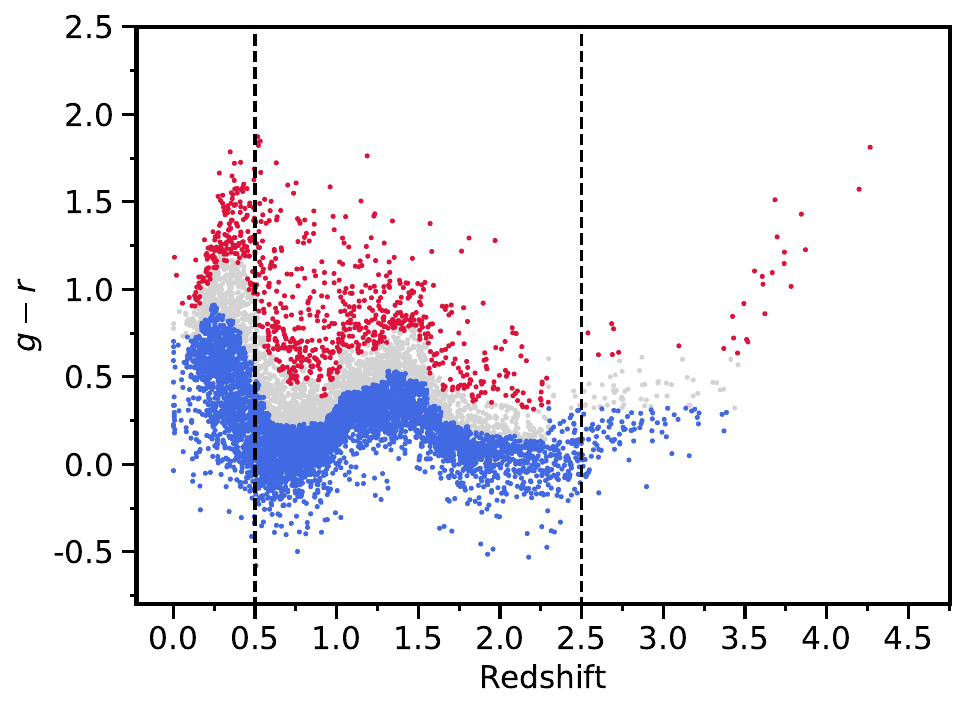}
    \caption{Redshift evolution of $g-r$ colour. rQSO are selected to be the 10$\%$ reddest sources in each redshift bin, whereas the blue quasars are the 2/3 bluest objects, to avoid contaminating both populations with sources close to the limit. Only sources within the two vertical dashed lines are kept in our sample ($z = 0.5 - 2.5$). }
    \label{fig:gr_z}
\end{figure}

To separate red quasars from the broader blue quasar population, we adopt a redshift-dependent photometric-selection based on their observed $g-r$ colour, similar to the method described in \citealt{fawcett_fundamental_2020,fawcett_fundamental_2022}. We split our parent X-ray--selected QSO sample into bins of redshift with a fixed number count of $N=200$ sources (which divides our parent sample into 40 bins of redshift), and select the $10\%$ reddest in each bin as our red quasar population. This selection method, based on observed-frame colour, is a relative and observational definition. To minimize contamination of both populations with sources lying close to the separation, we set a buffer region in between both populations, and only select the bluest 2/3 quasars in each bin to separate the parent population of bQSOs. Figure \ref{fig:gr_z} shows the distributions of sources in the $(g-r)$--redshift space, highlighting the selected rQSO and bQSO samples. Throughout the rest of the paper, we only consider sources lying within the two dashed vertical lines, delimiting the $z=0.5 - 2.5$ region, to ensure spectroscopic coverage of the \ion{Mg}{II}\,$\lambda2800$ emission line at all redshifts, for BH mass measurements. We apply a signal--to--noise ratio cut on the quality of spectra: $S/N > 1.5$ per spectral pixel over the full wavelength range of each spectrum. Finally, we require our sources to have a $W1$ counterpart in the WISE catalogue for robust host characterization. Our final sample consists of 3763 control blue QSO, and 380 red QSO. We summarize in Table \ref{tab:ncount} the number of sources at each cut.

\subsection{Optical spectral fitting and reconstruction}
\begin{figure*}
    \centering
    \includegraphics[width=\linewidth]{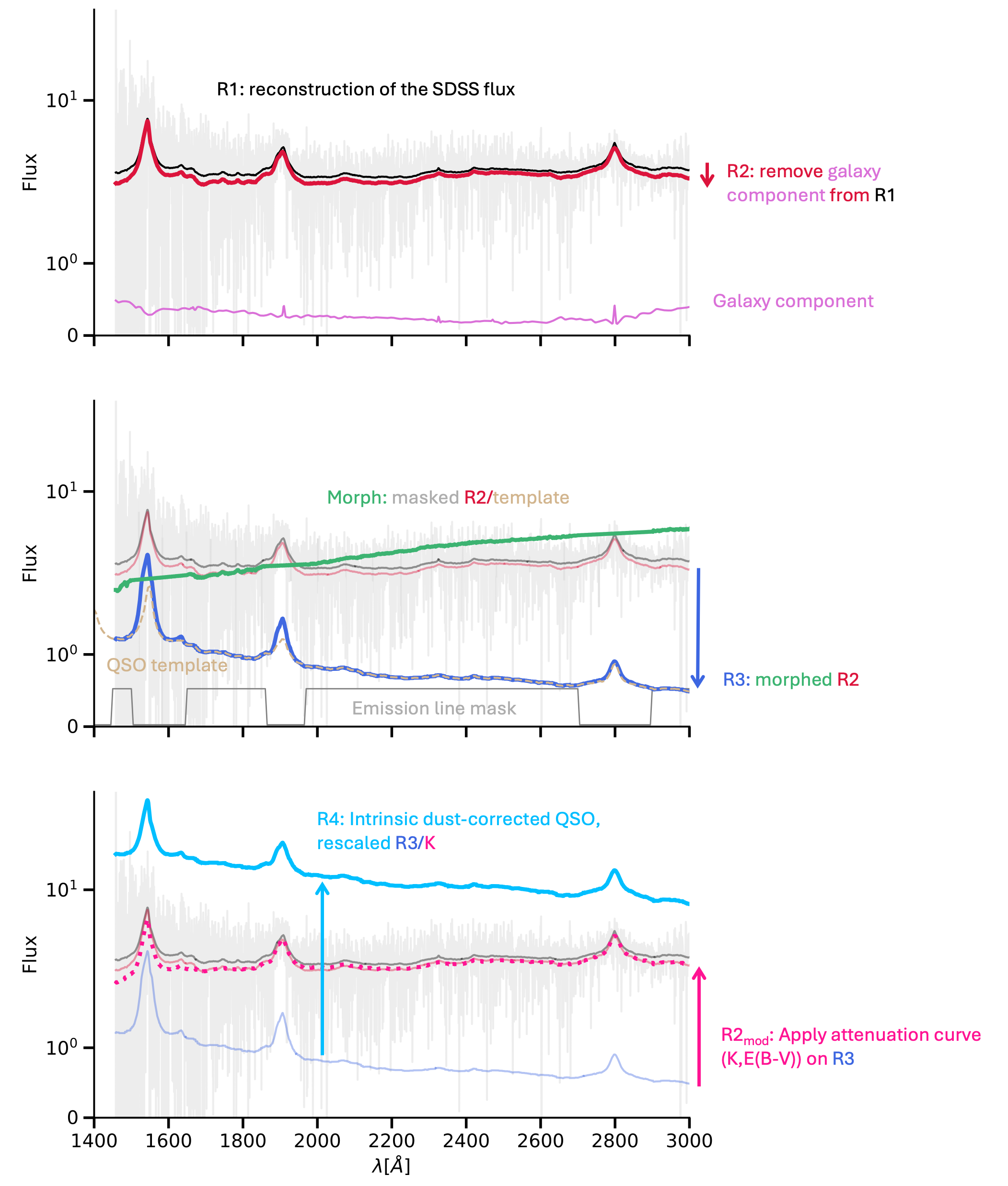}
    \caption{Diagram of the reconstruction process and reddening correction; see Section \ref{sec:specfit} for details. Colour coding is maintained across panels. \textit{Top panel:} We perform a first QSO+galaxy component reconstruction (R1 in black) of the total observed SDSS spectrum (light grey) and subtract the host-galaxy contribution (purple) to obtain a pure observed QSO reconstruction (R2 in red). \textit{Middle panel:} We create the morph array shown by the green line which is the ratio of the observed continuum and the QSO template continuum (dashed brown) with emission lines masked (mask shown in grey).  We divide R2 by the morph array to reshape the reconstruction and match it to the slope of the QSO template, which produces a de-reddened QSO reconstruction, R3 in blue. \textit{Bottom panel:} We derive a physical extinction $E(B-V)$ by applying an attenuation curve to R3 to best fit R2 (dashed pink). The median normalization $K$ is used to rescale R3 and produces R4 in light blue, the intrinsic dust-corrected QSO reconstruction.}
    \label{fig:sketch}
\end{figure*}

We now describe our method to fit models to the optical spectra of our quasar samples that allows the reconstruction of the \emph{intrinsic} underlying spectra and measure physical properties of both our rQSO and bQSO samples.
The first step involves reconstructing the global continuum shape and rough emission line morphologies by fitting a combination of independent additive components that characterise both the quasar and host-galaxy emission. This performs an initial characterisation which crucially allows an estimation and subtraction of the contribution to the optical spectrum from any underlying host galaxy (Section \ref{sec:ICA}). 
We then correct the reconstructed shape of the quasar light for reddening due to dust extinction (Section \ref{sec:reddening}), enabling robust estimates of both the intrinsic optical luminosity and the amount of dust along the line-of-sight to the quasar. 
Finally, we reconstruct the emission line profiles of the reshaped spectra to enable accurate measurements needed to derive physical properties, in particular black hole masses (Section \ref{sec:10c}). Figure \ref{fig:sketch} provides a visual summary of our process, while more details on each step are given in the section below.

\label{sec:specfit}
\subsubsection{Reconstructing QSO spectral shape and host decomposition}
\label{sec:ICA}
We perform spectral fitting using a non-parametric and data-driven approach to avoid making assumptions on spectral shape, as described in \citealt{rankine_bal_2020}. Rather than modelling spectra with analytical functions such as Gaussians for line profiles and power laws for continua, we reconstruct QSO spectra using a physically-motivated model, assuming that QSO spectra share a common underlying shape that can be reconstructed through a linear combination of components derived from a large sample of observed spectra. We adopt QSO components from \citealt{rankine_bal_2020} which were created using Mean-Field Independent Component Analysis (MFICA; \citealt{hojen-sorensen_mean-field_2002, allen_classification_2013}) applied to $\sim 4000$ individual QSO spectra from SDSS DR14 \citep{paris_sloan_2018}, covering the rest-frame wavelength range $1260 - 3000$ \r{A}. This range was chosen to cover emission lines relevant to the work in \citealt{rankine_bal_2020} across a broad range of redshift. Here we detail the process of spectral reconstruction using the QSO components from \citealt{rankine_bal_2020}, combined with five galaxy components from the \ascii{kcorrect} \ascii{python} module \citep{blanton_k_2007} generated from non-negative matrix factorization.

We perform a first reconstruction of the overall shape of our spectra and the emission lines (denoted $R1$, shown by the smooth black line on the top panel of Figure \ref{fig:sketch}) by fitting a linear recombination of the eight QSO component set and the five galaxy components. This set of eight QSO components is used to reconstruct the global continuum shape of our individual spectra as well as the broad morphology of the main emission lines. By combining both QSO and galaxy components, we are able to account for host-galaxy contribution, which primarily affects low-redshift, low-luminosity QSOs. The host-galaxy component is subtracted from the original flux, which provides $D2$, the QSO-only data (not shown on Figure \ref{fig:sketch} for clarity). We then subtract the host-galaxy component from the reconstruction $R1$ to derive a pure QSO smoothed spectrum ($R2$, red line). To recover the intrinsic, reddening-corrected spectral slope, we morph the observed reconstruction to match the shape of a typical unreddened QSO template, here using the template from \citealt{maddox_large_2012}, described in \citealt{rankine_bal_2020}. To achieve this reshaping, we compute a wavelength-dependent morph array (green line on the middle panel), defined as the ratio of the continua between the observed QSO-only reconstruction ($R2$) and the template outside of the masked emission lines, where the continuum is interpolated linearly. This morph array quantifies the deviation in spectral shape between the observed spectrum and the template due to extinction. We then divide the reconstruction $R2$ by this morph array, producing a third reconstruction, denoted $R3$ shown by the blue line, whose continuum has been morphed on to the template's continuum shape, capturing the overall intrinsic spectral shape, with emission line profiles. The host-subtracted data, \textit{D2}, is also morphed in a similar manner but is not shown in Figure \ref{fig:sketch} for clarity.

\subsubsection{Dust correction}
\label{sec:reddening}
As we selected the rQSO sample to have red photometric colours, we expect some amount of dust along the line of sight, reddening the continuum of the QSO component. A minor contribution to the observed reddening may arise from differential atmospheric refraction affecting the spectrophotometry, particularly at blue wavelengths. This effect can introduce small biases in the measured extinction for suboptimal spectrophotometric conditions. 

This dust along the line of sight makes our red sample appear somewhat different than the QSO template used for morphing, built to resemble a typical blue QSO. We can use this difference to directly measure the spectroscopic extinction in both populations, by quantifying the reddening needed to be applied to the morphed reconstruction $R3$ to best fit the observed QSO-only data $D2$, acting as a model of $R2$ by reddening $R3$.

We choose to model the optical extinction $E(B-V)$ and normalization $K$ using a powerlaw attenuation curve as motivated in \citealt{fawcett_fundamental_2022}, which describes the intrinsic flux as:

\begin{equation}
    F_{int}(\lambda) = KF_{obs}(\lambda) \times 10^{0.4 A_\lambda},
\end{equation}

\begin{equation}
    A_\lambda = R_V \times E(B-V) \left ( \frac{\lambda}{5500}\right )^\alpha,
\end{equation} 
with $R_V = 4, \alpha = -1$. We explore values of extinction $E(B-V)$ and normalization $K$ through MCMC with Stan \citep{Stan} using a Gaussian likelihood, by applying the attenuation curve to $R3$ until it best matches the slope of the galaxy-subtracted data $D2$, to measure reddening itnrinsic to the QSO. After iterating until the chains converge, we use the median $K$ and $E(B-V)$ from the posterior distributions of the individual sources to create $R2_{model}$ (pink dashed line in the bottom panel of Figure \ref{fig:sketch}), the best-fit of $D2$. To obtain the intrinsic de-reddened QSO-only reconstruction, indicated as R4 in light blue, we rescale the morphed reconstruction $R3$. Rescaling by the median $K$ recovers the true continuum level that was originally attenuated.

\subsubsection{Accurate reconstruction for emission line measurements}
\label{sec:10c}
Once the continuum shape has been reconstructed following Section \ref{sec:ICA} and corrected for dust as per Section \ref{sec:reddening}, we perform a final reconstruction on the morphed flux using the `standard' 10 QSO component set from \citealt{rankine_bal_2020} to reconstruct emission lines. This set of components is generated from the same training sample of 4,000 DR14 quasars but with their continuum shape morphed as described in Section~\ref{sec:ICA}. As described in section~2 of \citealt{rankine_bal_2020}, reconstructing the morphed spectra reduces the number of components necessary to achieve a specified accuracy. The `standard' set produces accurate reconstructions of the continuum and emission lines covering 1260--3000\,{\AA}, with $\sim$ 90$\%$ of spectra having a median fraction error ($(\rm data - reconstruction)/reconstruction$) of less than 2$\%$.

\subsection{SED fitting}
\label{sec:SED}

To derive the physical properties of host-galaxies, such as stellar masses, we employ Spectral Energy Distribution (SED) fitting, using the extensive photometric data detailed in Section \ref{sec:photometry}. The reliability of stellar mass measurements is however challenged by the presence of AGN emission. This can introduce scatter in the $SFR - M_\star$ relation, due to uncertainties driven mostly by stellar mass measurements, which are often overestimated due to AGN contribution \citep{ciesla_constraining_2015,buchner_genuine_2024}. However, our comprehensive wavelength coverage allows us to compare SEDs between the bQSOs and rQSOs fairly across a wide range of redshifts and constrain the contributions from different processes including stars, dust, and active galactic nuclei (AGN).

Many codes are publicly available to perform SED fitting. Here, we chose to use the \textsc{python} Code Investigating GALaxy Emission \citep[\textsc{cigale};][]{boquien_cigale_2019}, making use of its short computing time and low disk storage needs as well as innovative additional AGN and X-ray components in the model of SEDs \citep{yang_fitting_2022}. In the following, we describe the models we built and the components used to accurately fit the SEDs of quasar-dominated galaxies.

We adopt the same grid of models to describe galaxy and AGN contributions to the SED of both the rQSO and control samples. To incorporate the temporal evolution of star formation rates (SFRs), we choose a delayed Star Formation History (SFH) model. This model offers smoother variations in SFR over time rather than the sudden onset of star formation that is assumed when modelling galaxies with exponentially declining SFH models. 
It reaches a peak at $\tau_{main}$, then gradually decreases.

To describe the stellar population, we adopt the widely used library defined in \citealt{bruzual_stellar_2003} using a grid of fixed metallicity along with the Salpeter Initial Mass Function (IMF). To differentiate between young and old stars, we allow for an age-dependent reddening factor, to account for young stellar populations still embedded in their dust clouds which absorb at short wavelengths, in addition to absorption by dust in the Interstellar Medium (ISM). To account for this age dependency, we assume an attenuation law from \citealt{charlot_simple_2000} which allows the computation of two different attenuation contributions: dust clouds surrounding young stars and the ISM for stars of all ages. Additionally, we account for nebular emission manifested by emission lines from gas ionised by the radiation field of young stars and absorption by the intergalactic medium (IGM) using the model developed by \citealt{inoue_updated_2014}. This model considers the emission originating from the most massive stars, which ionize the surrounding gas, resulting in the re-emission of light through a series of emission lines. Dust emission is modelled as in \citealt{draine_andromedas_2013}, and is separated into two components. The first component is the diffuse dust emission, heated by the global stellar population, while the second component models dust emission specifically linked to star-forming regions, where, contrary to the first case, dust is heated by a variable radiation field ranging from $U_{min}$ to $U_{max} = \rm 10^7$. 

We add an AGN component following the SKIRTOR clumpy torus model \citep{stalevski_3d_2012,stalevski_dust_2016} to disentangle the nuclear emission from star formation; both contributing strongly in the UV. AGN emission is modelled using three radiative components: the primary source located within the torus, the scattered emission by dust, and thermal dust emission \citep{casey_far-infrared_2012}. The attenuation by dust is modelled by the empirical SMC curve, which resembles our analytical model using a power-law extinction curve described in Section \ref{sec:reddening}, but starts slightly deviating below $\lambda \sim 2500$ \r{A}. A set of several parameters describes this component, including the radius $r$ of the torus and the fraction of light coming from the AGN.

Finally, we include X-ray fluxes in the total SEDs, modelled as a galaxy component to include the minor contribution from hot stars and off-nucleus binary systems, and an AGN component.

Full details of the parameter grid used can be found in Table~\ref{cigale_param}.

\section{Results}
\label{sec:results}
We now present our results on the differences in intrinsic properties between rQSO and a bQSO control sample defined in Section \ref{sec:selection}. Section \ref{sec:LX} examines their differences in terms of X-ray luminosity as calculated in Section \ref{sec:erosita}, to establish a first comparison of their accretion properties. Section \ref{sec:spec} derives spectral properties of red and blue quasars after careful dust and host correction following the method outlined in Section \ref{sec:specfit}, to measure optical extinction by dust, single-epoch virial BH masses, and Eddington ratios. Finally, we investigate the host-galaxy properties of rQSO in Section \ref{sec:host} to assess whether they could also differ at the galaxy scale.

\subsection{X-ray properties of reddened Quasars}
\label{sec:LX}
\begin{figure}
    \centering
    \includegraphics[scale=0.55]{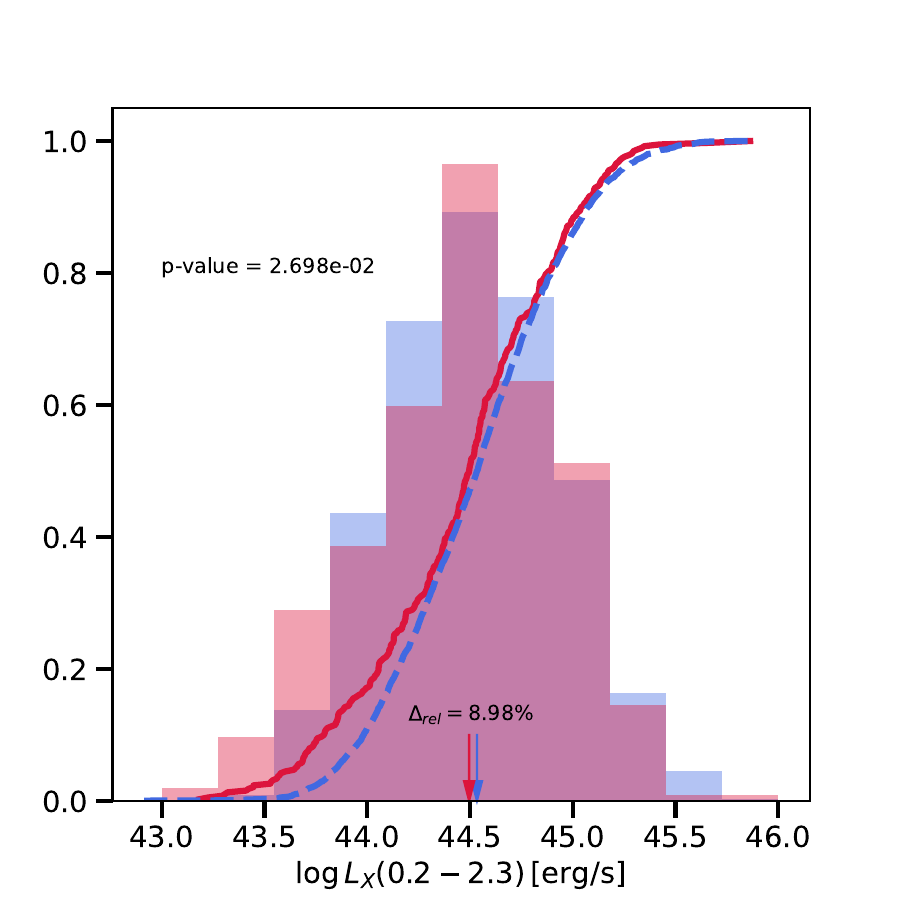}
    \caption{X-ray luminosity distribution of red and blue quasars photometrically defined (see Section \ref{sec:selection}). The Cumulative Distribution Functions (CDF) displayed as the solid and dashed lines to compare both populations without binning effects. There is an over-density of rQSOs at the low-end of the luminosity distribution compared to bQSOs. We quantify the statistical significance of this difference between the two CDFs through the two-sample KS-test, giving a $p$-value $<$ null-hypothesis 0.05, which is evidence for a significant difference between the two samples.}
    \label{fig:LX}
\end{figure}

\begin{figure}
    \centering
    \includegraphics[scale=0.55]{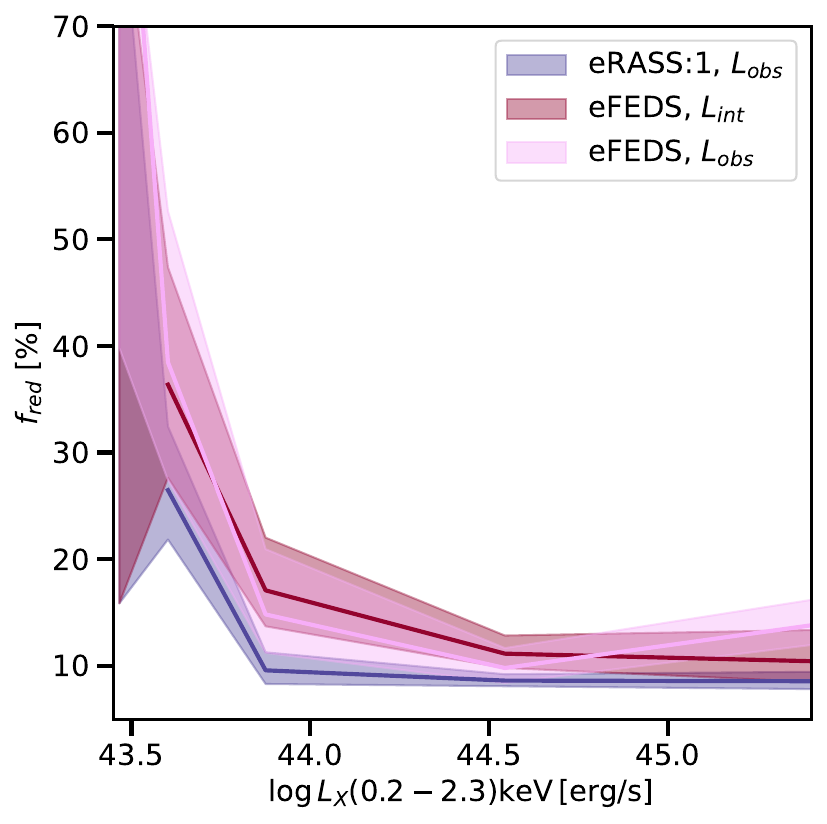}
    \caption{Fraction of rQSOs as a function of various X-ray luminosities: observed $L_X$ from our full sample within eRASS1 (light pink), intrinsic $L_X$ corrected for obscuration in eFEDS (fuchsia) and observed $L_X$ from our sample within eFEDS only (dark red). All relations show a decrease of the fraction of rQSO with increasing luminosity regardless of obscuration effect by dust and gas.}
    \label{fig:fred}
\end{figure}

We investigate here the potential differences in the X-ray emission of red and blue quasars and their prevalence as a function of X-ray luminosity.

Figure \ref{fig:LX} displays the distribution of observed X-ray luminosity in the 0.2$-$2.3 keV band for both samples, red and blue. Both distributions span roughly the same range in luminosity, covering $L_X \sim 10^{43-46} \, \rm erg\, s^{-1}$, as well as having a similar mean, with both distributions peaking around $L_X = 10^{44.5} \, \rm erg \,s^{-1}$. Despite those similarities, there is an over-density of rQSO in the low-end of the luminosity distribution ($L_X < 10^{44} \, \rm erg \, s^{-1}$) compared to the control sample. To assess the significance of this difference, we turn to the Cumulative Distribution Function (CDF) indicated by the solid lines in Figure \ref{fig:LX}. Directly examining the CDF removes any biases induced by the binning effect of a standard distribution. Here, the relative over-density of rQSOs at low X-ray luminosity is even clearer, as the red CDF dominates at the faint end, which is then over-taken by the blue CDF at the brighter end. We then perform a 1-dimensional 2-sample KS test \citep{hodges_significance_1958} which compares the CDF to determine where the maximum difference lies between the two distributions and evaluate its statistical significance. We require a $p-$value lower than 0.05 to reject the null hypothesis that the two distributions are identical at moderate significance ($\sim 2 \sigma$ equivalent). This test yields a $p$-value of $\sim 0.027$, lower than the $2\sigma$ threshold to reject the null hypothesis that the two samples were drawn from the same underlying distribution. This result suggests that while the magnitude of the differences seems to be moderate, the red and blue QSO samples are statistically different in terms of their X-ray emission, and that the distribution of rQSOs is more skewed towards lower X-ray luminosities than for the bQSO sample.

To determine the relative abundance of rQSO as a function of X-ray luminosity, we compute $f_{red}$, the fraction of rQSO relative to the total number of QSOs in the parent sample in bins of $L_X$, shown in Figure \ref{fig:fred}. We observe a clear decrease in $f_{red}$ with increasing X-ray luminosity, reflecting the higher prevalence of rQSO at lower $L_X$ already identified. However, we stress that the X-ray luminosities derived in this work for our full sample from eRASS1 have not been corrected for intrinsic obscuration, only Galactic absorption (see Section \ref{sec:erosita}). As such, the dark blue curve showing the $f_{red} - L_X$ relation across our full eRASS1 sample is based on observed luminosities ($L_{obs}$) that may be underestimated due to intrinsic absorption effects. To ensure that the prevalence of rQSO at faint X-ray luminosities is not driven by obscuration effects causing an under-estimate of the true luminosity of dusty quasars, we use the intrinsic luminosities ($L_{int}$) of a sub-sample from eFEDS \citep{liu_erosita_2022}, which are corrected for absorption based on X-ray spectral fitting, providing measurements of the absorbing column density $N_H$ along the line-of-sight. We only retain sources with sufficient fit quality as explained in Section \ref{sec:obscuration}. This sample contains a total of 720 bQSOs and 134 rQSOs. The curves in light pink and dark red in Figure \ref{fig:fred} show the $f_{red} - L_X$ relation in eFEDS using intrinsic (absorption-corrected) and observed X-ray luminosities, respectively. Regardless of the absorption correction, the same trend remains for all three cases, where rQSOs are more prevalent at fainter X-ray luminosity. This suggests that rQSOs are intrinsically X-ray weaker than bQSOs.
However, obscuration and absorption of the X-ray emission by dust could still play a significant part in the difference between observed and intrinsic properties, especially at soft energies where luminosities are more sensitive to absorption and where eROSITA is the most sensitive. We note that while absorption correction based on X-ray spectral fitting was done on the eFEDS sample, some sources have unconstrained column density measurements, especially at faint luminosities and are removed from our sample. The sources kept in our analysis have on average over 20 counts per source, both rQSOs and bQSOs, satisfying a high enough photon count threshold for reliable column density measurements. We discuss these limitations and potential effect of absorption on our results in Section \ref{sec:obscuration}.

\begin{figure*}
    \centering
    \includegraphics[width=\linewidth]{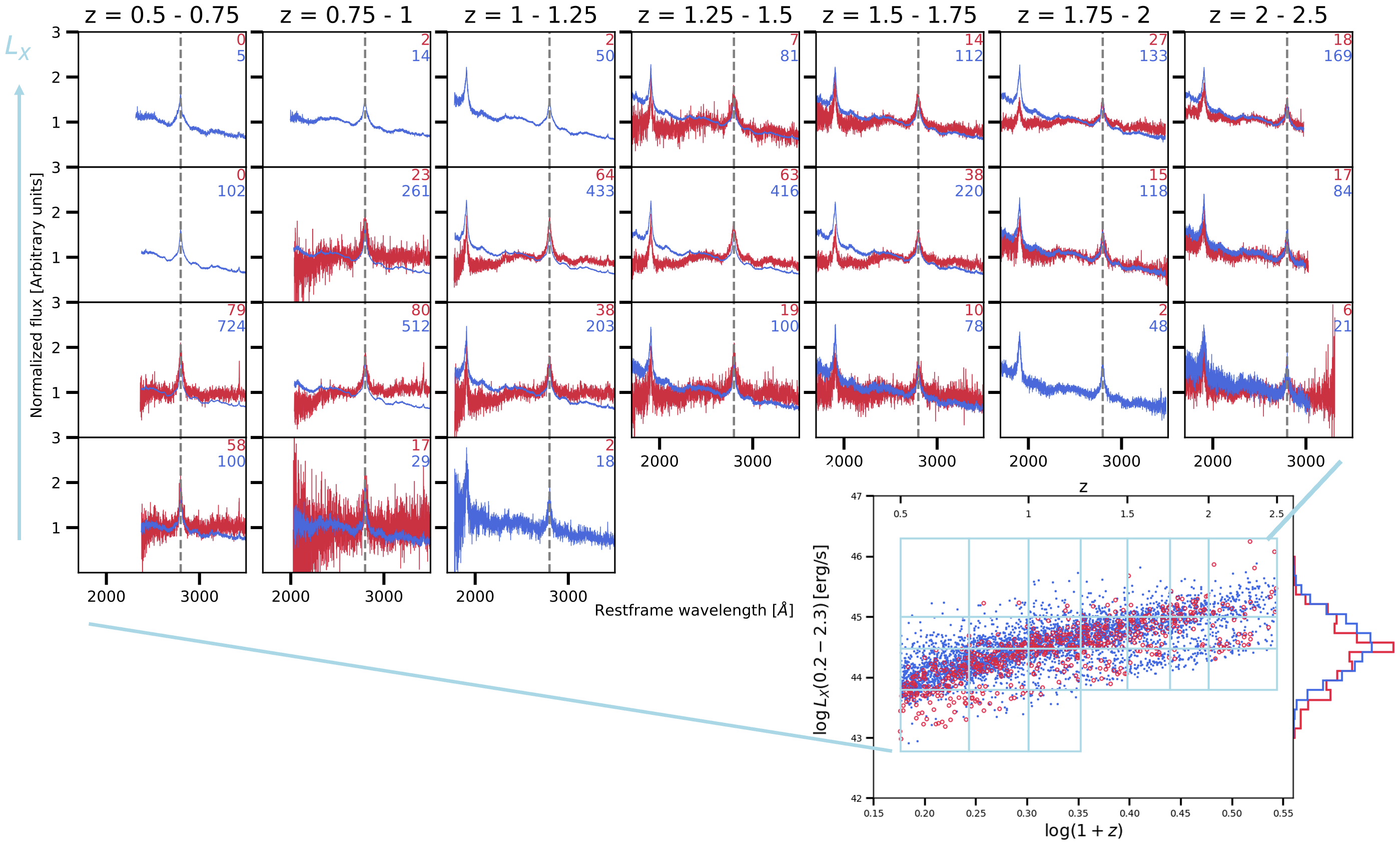}
    \caption{\textit{Bottom:} Distribution of rQSOs (red open circles) and the control blue sample in the X-ray luminosity - redshift space. No difference is found in the redshift distribution of rQSO as their selection is $z$-dependent. Both luminosity distributions peak around $L_X \sim 3\times 10^{44} \, \rm erg \, s^{-1}$ but with an over-density of rQSOs in the low-end of the distribution. \textit{Top:} We grid this 2D space in bins of $L_X - z$ and produce composite spectra in each bin, showing the luminosity-redshift evolution of the spectral shape of rQSOs and the control sample.}
    \label{fig:gridspec}
\end{figure*}

\subsection{Spectroscopic properties of reddened Quasars}
\label{sec:spec}
Here, we derive spectral properties of red and blue quasars after careful host decomposition and dust correction as described in Section \ref{sec:specfit}. After quantifying optical extinction by dust in Section \ref{sec:ebv}, we determine whether rQSOs show intrinsic differences with bQSOs in terms in terms of accretion properties by deriving Eddington ratios (Section \ref{sec:edd}).

In Figure \ref{fig:gridspec}, we split our sample in bins of $L_X-z$ (\textit{bottom panel}) and create composite optical spectra for the rQSOs and bQSOs in each bin. To generate the composite spectra, we first normalise each individual spectrum over the 2575-2625\AA ~wavelength range (where the continuum dominates). We then take the median of the normalized spectra over their common wavelength range to produce one reference spectrum per $L_X - z$ bin for both the rQSO and bQSO samples (\textit{top panel}). The numbers displayed on the top right corner of each cell is the number of spectra we average over to produce one single composite. We require a minimum of five individual spectra in a bin to generate a meaningful composite. The dashed grey vertical line identifies the \ion{Mg}{II} emission line at 2800\AA, which is always spectroscopically covered in our selected redshift range $z = 0.5 - 2.5$. From the composite spectra, we see that our control bQSO sample conserves the blue spectral slope such that the emission drops toward redder wavelengths, the typical feature of QSOs, even at lower X-ray luminosities, where we may expect more obscured AGN. With our selection, we can thus confidently confront our rQSO sample with the bQSOs acting as reference for normal unreddened QSOs. As a first look at the spectroscopic properties of red and blue QSOs, the composite spectra show striking differences in the continuum shape. The flux in the composite of rQSO is significantly attenuated in the blue end of the wavelength range, making them appear much less steep than the control sample. We observe a redshift evolution of this slope in rQSO at fixed X-ray luminosity (moving to the right across the grid), where their continuum slope is more similar to bQSOs with increasing redshift. This result shows that rQSOs might be less affected by dust attenuation at higher redshift, which is most likely a selection effect from our $S/N$ requirement and our photometric definition of rQSO (see Section \ref{sec:selection}) struggling to detect strongly reddened broad-line AGN as seen in Figure \ref{fig:gr_z}. This could be due to the wider high-z bins which includes moderate $g-r$ excess that would have been classified as bQSO at lower redshift. However, Section \ref{sec:ebv} introduces a more physical and independent way to select rQSOs based on their inferred spectral extinction $E(B-V)$, showing that our results do not rely exclusively on a photometric selection.

While Figure \ref{fig:gridspec} serves as illustrative purposes, we now individually measure in detail spectroscopic properties of rQSO and their X-ray luminosity and redshift dependences from the individual spectra directly and compare to a $L_X-z$ matched control sample of bQSOs.

\subsubsection{Optical extinction}
\label{sec:ebv}
We model the optical extinction $E(B-V)$ using a power-law attenuation curve as described in Section \ref{sec:reddening}. We perform a Bayesian fit on each individual source with the same exploration range for extinction and scale parameters $0\leqslant E(B-V) \leqslant 2.5 \, , 0\leqslant K \leqslant 10^3$, with $K$ the parameter rescaling $R3$ to the intrinsic flux level of the dust-corrected QSO-only reconstruction. We use the same uniform prior for both samples of red and blue quasars. For each source, the best estimate of $E(B-V)$ is taken as the median of the posterior distribution once the fit is completed. 

\begin{figure}
    \centering
    \includegraphics[scale=0.5]{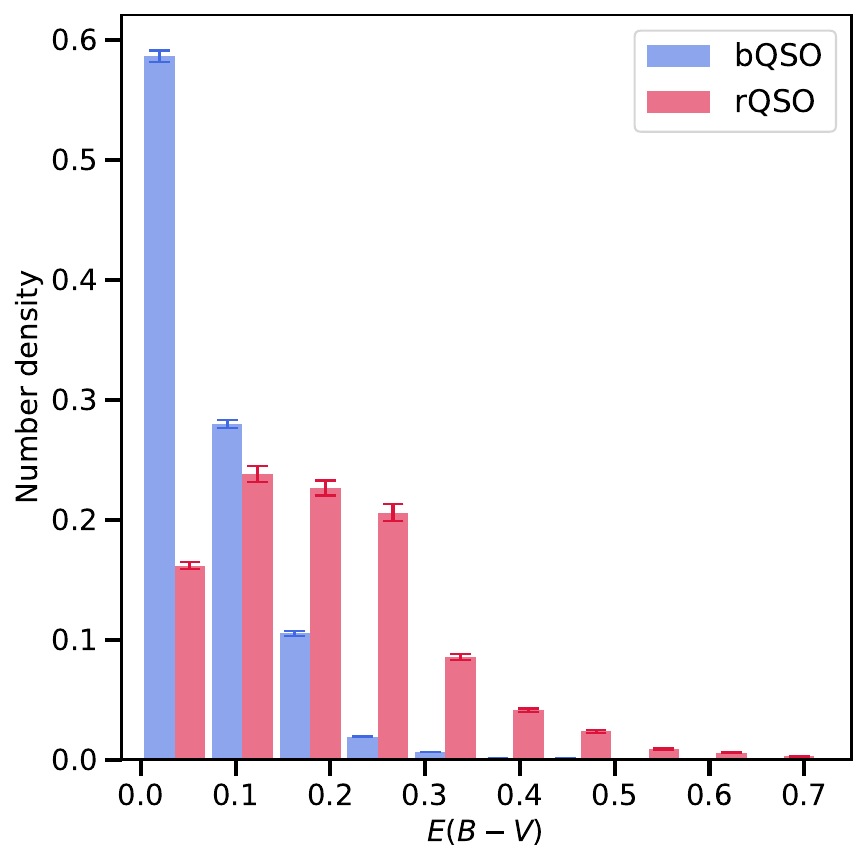}
    \caption{Distribution of the median of the posterior distributions of the optical extinction $E(B-V)$ for red and blue quasars. The error bars are the average normalized 1$\sigma$ spread in each bin.}
    \label{fig:ebv}
\end{figure}

Figure \ref{fig:ebv} displays the 1-dimensional distribution of the measured extinction, $E(B-V)$. The error bars represent the average normalized 1$\sigma$ spread in each $E(B-V)$ bin. The distributions for rQSO and bQSO show many differences: rQSO span a much broader extinction range than bQSO with $E(B-V)_{red} < 0.7$, against $E(B-V)_{blue} < 0.35$. The median of the extinction distribution of rQSO is $\langle E(B-V)\rangle_{red} = 0.21$ and $\sim0.04$ for bQSO. The two distributions have little overlap at high $E(B-V)$ values because of the wide spread of rQSO across extinction values and the rapid decline in the bQSO distribution. This result shows that from an observed photometric selection of rQSO, we can robustly recover a sample of optically attenuated quasars due to dust along the line-of-sight measured spectroscopically. Our method for spectroscopic extinction measurement can provide another robust definition of the rQSO sub-sample, selected as the high-end $E(B-V)$ tail of the distribution. Cutting at $\langle E(B-V)\rangle>0.35$, where the bQSO distribution stops and no overlap remains between the two distributions, would provide a clean and robust sample of rQSO, conserving over $10\%$ of the original sample of rQSOs. For a less conservative definition, merging the two distributions and only selecting QSO with $\langle E(B-V)\rangle>0.21$, the median of the current rQSO distribution, would retain $2\%$ of the total bQSO sample, which are QSOs with high spectroscopic extinction but missed in our selection due to their blue photometric colours. We note that the interpretation of individual extinction measurements should be treated with caution due to possible host contribution residuals, especially in the faintest or reddest systems. However, our interpretation relies on robust statistical trends across our full sample and our conclusions remain.

\begin{figure}
    \centering
    \includegraphics[scale=0.55]{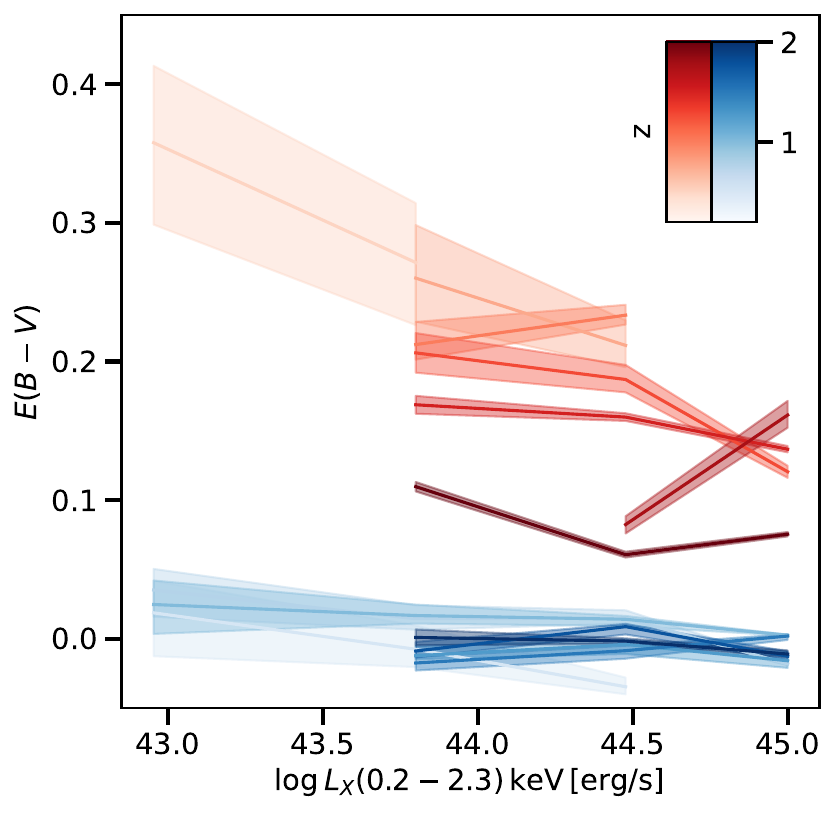}
    \caption{Evolution of the intrinsic optical extinction $E(B-V)$ as a function of X-ray luminosity and redshift for rQSO and bQSO. The solid lines are the median $\langle E(B-V)\rangle$ and the shaded areas show the 1$\sigma$ spread in bins of $L_X-z$. Redshift increases with darker shades. In both samples, we find a larger average extinction at faint X-ray luminosities and a decrease in $E(B-V)$ with increasing $L_X$. bQSOs show on average very little extinction even in the faintest quasars, and do not overlap with rQSOs in terms of $E(B-V)$ ($<0.1$).}
    \label{fig:ebv_lx}
\end{figure}

Figure \ref{fig:ebv_lx} presents the X-ray luminosity and redshift dependence of the optical extinction $E(B-V)$ for rQSOs and bQSOs. The solid lines are the median $\langle E(B-V)\rangle$ and the shaded areas show the 1$\sigma$ spread in bins of $L_X-z$. Redshift increases with darker shades. For both samples, there is a decreasing trend of $E(B-V)$ with increasing luminosity, indicating that the intrinsic emission from X-ray faint quasars is more likely to be attenuated by dust. Again, we see clearly here that the red and blue quasar samples do not overlap in terms of extinction, and this is true at all redshifts. Indeed, bQSOs span a small range of $E(B-V) \sim 0-0.05$ on average, with very little $L_X-$evolution, whereas rQSOs span a much broader range of optical extinction and at higher values than bQSO with $E(B-V) \sim 0.1 - 0.4$ on average. Examining the dependence of the extinction with redshift at fixed luminosity, $E(B-V)$ appears to decrease with increasing redshift for rQSO. This behaviour could represent an evolutionary effect that quasars become progressively enshrouded by dust with time. However, there could be strong selection effects biasing this correlation due to our redshift-dependent definition of rQSO.

\subsubsection{Eddington ratio}
\label{sec:edd}
We now assess whether intrinsic differences between rQSO and bQSO are found in their accretion properties rather than mass. To trace instantaneous accretion, we derive Eddington ratios $\lambda_{Edd}$ from their intrinsic UV luminosities (corrected for dust extinction and host contamination following Section \ref{sec:specfit}) such that:

\begin{equation}
    \lambda_{Edd} = \frac{L_{Bol}}{L_{Edd}} = \frac{k_{Bol}L_{3000}}{1.26\times 10^{38}M_{BH}}
\end{equation}

BH masses ($M_{BH}$) are derived from single-epoch virial BH masses using \ion{Mg}{II} line and re discussed in Section \ref{sec:frac_growth}. Here we use a luminosity-dependent bolometric correction from \citealt{netzer_bolometric_2019}:

\begin{equation}
    k_{bol} = c\left(\frac{L_\lambda}{10^{42} \rm erg \, s^{-1}}\right)^d
\end{equation}
\noindent with $c,d = 25,-0.2$ for $L_\lambda = L_{3000}$.

\begin{figure}
    \centering
    \includegraphics[scale=0.5]{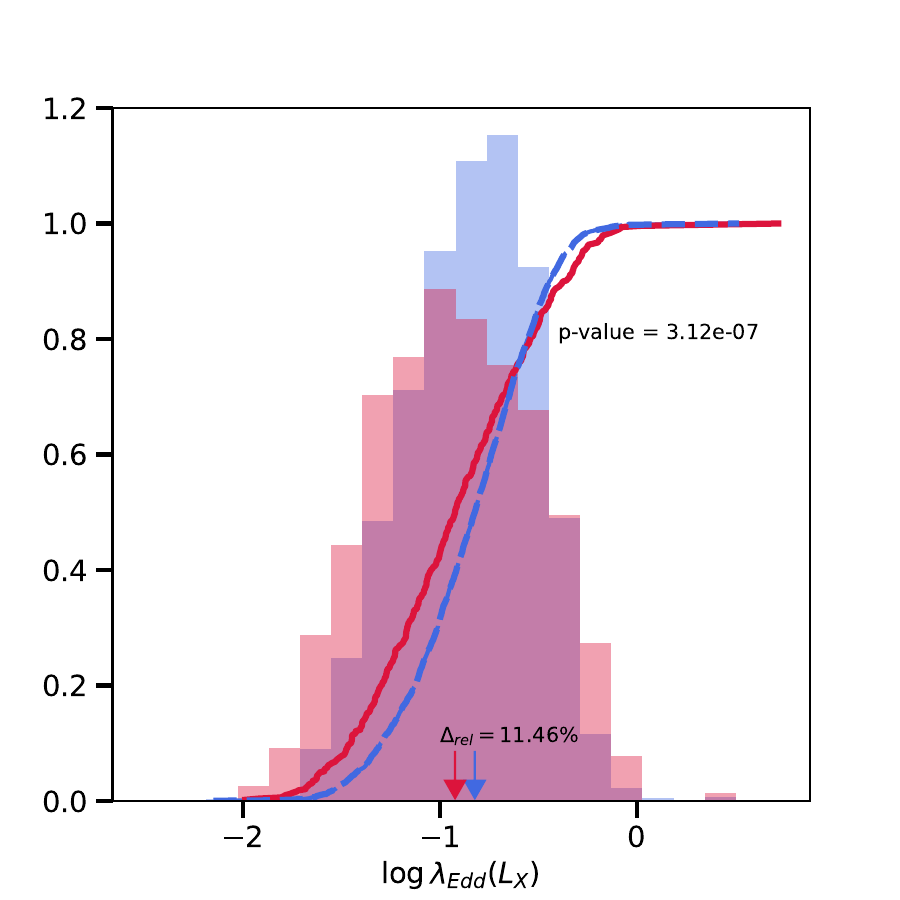}
    \caption{Distribution of the Eddington ratio measured from intrinsic UV luminosities. The median of the distributions are shown by the arrows and their relative difference $\Delta_{rel}$ is around 11.46$\%$. The Cumulative Distribution Functions (CDFs) are shown by the solid red and dashed blue lines. We perform the 2-sample KS-test on the CDFs, providing a $p$-value of $3.12\times 10^{-7}$.}
    \label{fig:edddistrib}
\end{figure}

Figure \ref{fig:edddistrib} displays the distribution of Eddington ratio for both samples. Both distributions span a wide range of $\log \lambda_{\rm Edd} = -2 - 0$, but it should be noted that the highest values $\log \lambda \gtrapprox 0.1$ could be driven by unreliable BH mass measurements where the virial assumption likely fails. Similar to Figures \ref{fig:LX} and \ref{fig:mbh-mstar}, we add the two CDFs and the medians of both distributions shown by the arrows. This time, the differences in $\lambda_{Edd}$ between red and blue quasars are much more statistically significant compared to the differences in $M_{BH}$, an offset that also persists between distributions at fixed BH mass, demonstrating that uncertainties and trends on $M_{\rm BH}$ are not driving the effect observed in $\lambda_{\rm Edd}$. We find a negligible $p$-value of $3.12\times 10^{-7} \ll 0.05$, and a larger relative difference of $\Delta_{rel} = 11.46\%$. This result shows that the difference in observed properties chosen to select red and blue quasars are driven by intrinsic effects, linked to differences in the accretion properties of BHs, and that rQSOs show systematically suppressed accretion relative to blue quasars.

\subsection{Relative optical and X-ray strength}
\label{sec:aox}
In section \ref{sec:LX}, we demonstrated that rQSOs are more prevalent in the low X-ray luminosity regime and that their absolute X-ray luminosity is, on average, lower than bQSOs. We now investigate this potential X-ray weakness relative to their optical emission by deriving the spectral slope indicator $\alpha_{OX}$ \citep{tananbaum_x-ray_1979}:

\begin{equation}
    \alpha_{OX} = \log\left(\frac{L_{2\rm keV}/L_{2500}}{\nu_{2\rm keV}/\nu_{2500}}\right) = 0.384 \log \left(\frac{L_{2\rm keV}}{L_{2500}}\right).
\end{equation}

The optical monochromatic luminosity $L_{2500}$ is taken as the dust-corrected continuum luminosity averaged over 2495-2505 \AA ~, divided by the 2500 \AA ~frequency. The X-ray monochromatic luminosity $L_{2\rm keV}$ is computed from the rest-frame 0.2-2.3 keV luminosity (see Section \ref{sec:erosita}), not corrected for obscuration since the effect of gas absorption is minimal on the X-ray emission as shown in Section \ref{sec:LX} and further discussed in Section \ref{sec:obscuration}. Extrapolating the rest-frame 0.2-2.3 keV luminosity to the 2-10 keV band, integrating over this energy range, and dividing by the 2keV frequency, this gives:

\begin{equation}
    L_{2\rm keV} = \frac{L_{2-10 \rm keV}}{\sum_{\nu_{2\rm keV}}^{\nu_{10\rm keV}} \nu^{1-\Gamma}\nu}\nu_{2\rm keV}^{\left(1-\Gamma\right)}
\end{equation}

\begin{figure}
    \centering
    \includegraphics[width=1.1\linewidth]{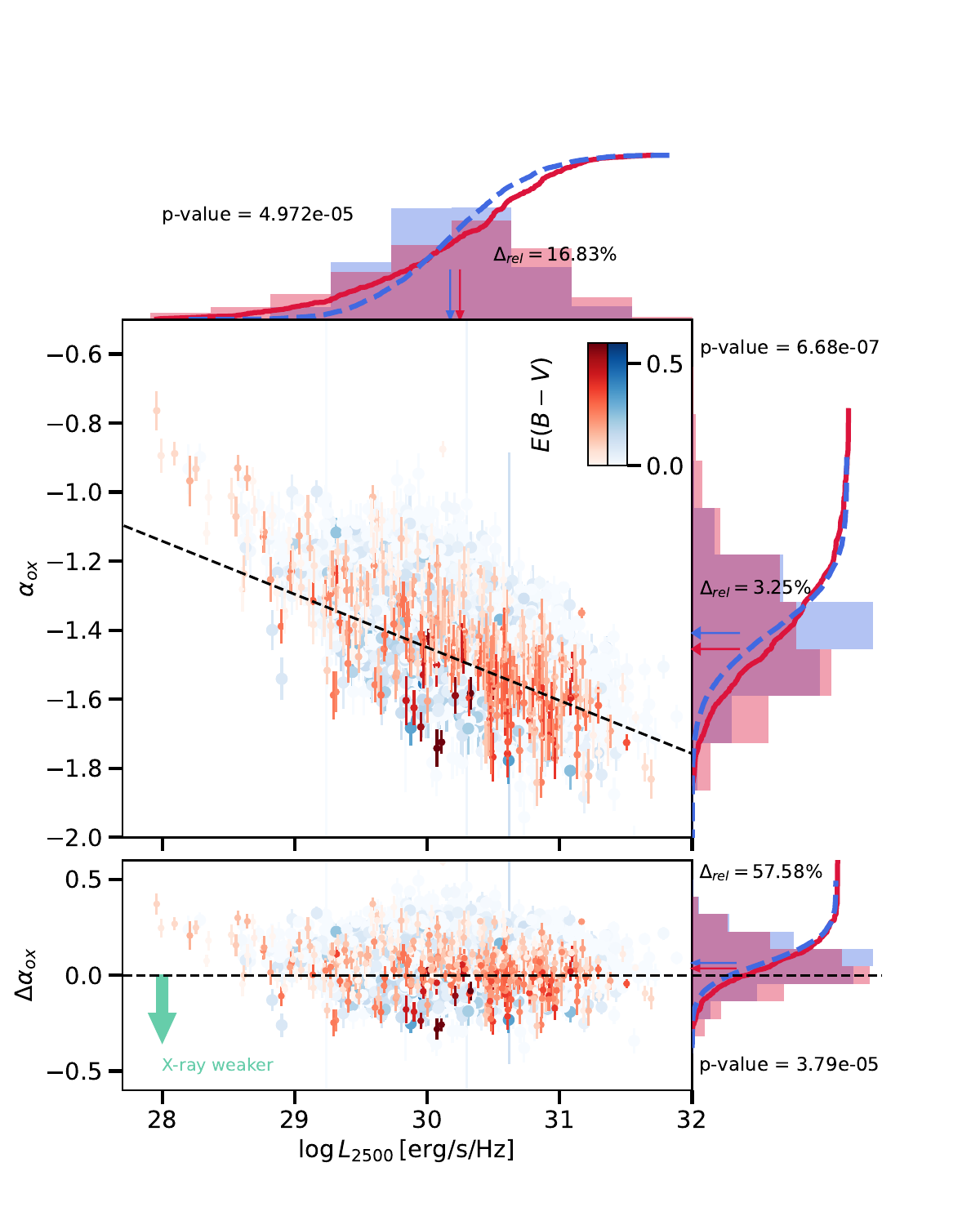}
    \caption{\textit{Center:} Distribution of blue and red quasars in the $\alpha _{OX} - L_{2500}$ space. The error bars show the uncertainties propagated from the 0.2-2.3 keV \textit{eROSITA} flux. The scaling relation indicated by the dashed black line is adapted from \citealt{lusso_x-ray_2010}. Points are colour-coded by their measured $E(B-V)$. A larger absolute value of $\alpha _{OX}$ (below the scaling relation) indicates weaker X-ray emission relative to optical. The side histogram is the $\alpha _{OX}$ distribution. The rQSO distribution is shifted towards lower $\alpha _{OX}$ relative to bQSO, indicating a stronger X-ray weakness relative to optical luminosity. The median of the distributions are shown by the arrows and their relative difference $\Delta_{rel}$ is around 3.25$\%$. The top panel displays the distributions of monochromatic luminosity $L_{2500}$ for red and blue quasars. rQSOs have a larger average $L_{2500}$ compared to bQSOs and a relative difference $\Delta_{rel} \sim 16.8\%$. The 2-sample KS-test yields a p-value of 4.972$\times 10^{-5}$. \textit{Bottom:} $\Delta \alpha _{OX}$, the distance of red and blue quasars to the \citealt{lusso_x-ray_2010} scaling relation, as a function of monochromatic optical luminosity $L_{2500}$. The side histograms show that rQSOs are more often X-ray weak than bQSOs.}
    \label{fig:aox}
\end{figure}

Figure \ref{fig:aox} presents the $\alpha_{OX} - L_{2500}$ relation (main panel) and compares our sample to the \citealt{lusso_x-ray_2010} scaling relation (dashed black line). The error bars are the uncertainties on the \textit{eROSITA} 0.2-2.3 keV flux propagated to the 2 keV monochromatic luminosity $L_{2\rm keV}$. We assume that the uncertainties on $\alpha_{OX}$ are dominated by uncertainties on the X-ray measurements and that the errors on optical luminosities are relatively negligible.

Both rQSOs and bQSOs follow the scaling relation from \citealt{lusso_x-ray_2010} with little to no offset. However, there is a significant shift in the $\alpha_{OX}$ distribution between the two populations (side panel, $p$-value $<0.05$), with rQSOs occupying systematically lower values of $\alpha_{OX}$ at fixed optical luminosity compared to bQSOs, indicating weaker X-ray emission compared to $L_{2500}$. At the same time, the 1D distribution of $L_{2500}$ shows that rQSOs are not simply faint objects across all bands, rQSOs are on average slightly more luminous in the optical than bQSOs, arguing against rQSOs being globally faint sources but rather have distinct X-ray weakness. The slight deviation at low optical luminosities from the scaling relation for both populations is due to our sample being X-ray selected and thus biased towards slightly brighter X-ray sources than the sample used to calibrate the relation. 

The bottom panel shows the distribution of $\Delta\alpha_{OX}$, defined as the deviation of sources from the \citealt{lusso_x-ray_2010} scaling relation. We observe a significant higher prevalence of rQSOs within the negative regime compared to bQSOs, with a higher fraction of sources classified as X-ray weak relative to predictions based on their optical luminosity, which reinforces that rQSOs are systematically X-ray weaker than bQSOs at matched $L_{2500}$.

Points are colour-coded by optical reddening $E(B-V)$. This format reveals a clear trend in both populations, mostly visible in rQSOs: at fixed optical luminosity, sources with higher extinction have lower $\alpha_{OX}$ and more negative $\Delta\alpha_{OX}$. This result indicates that larger dust content along the line-of-sight is associated with more reduced intrinsic X-ray emission relative to the optical, suggesting a link between the origin of the dust and X-ray weakness, which we discuss further in Section \ref{sec:discussion}.

\subsection{Host properties of reddened Quasars}
\label{sec:host}
\begin{figure}
    \centering
    \includegraphics[scale=0.5]{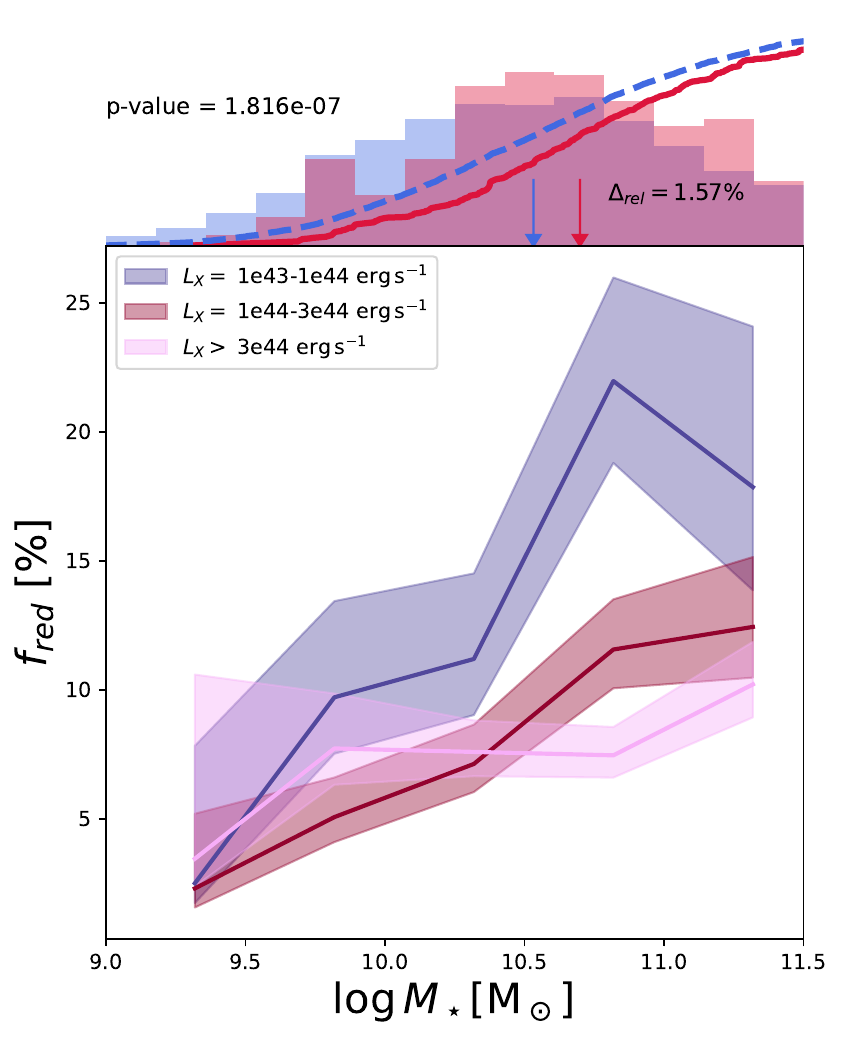}
    \caption{Fraction of rQSO, $f_{red}$, as a function of host stellar mass in different X-ray luminosity bins. The top panel displays the stellar mass distributions and CDFs for both populations.}
    \label{fig:fred_mstar}
\end{figure}

\begin{figure*}
    \centering
    \includegraphics[scale=0.65]{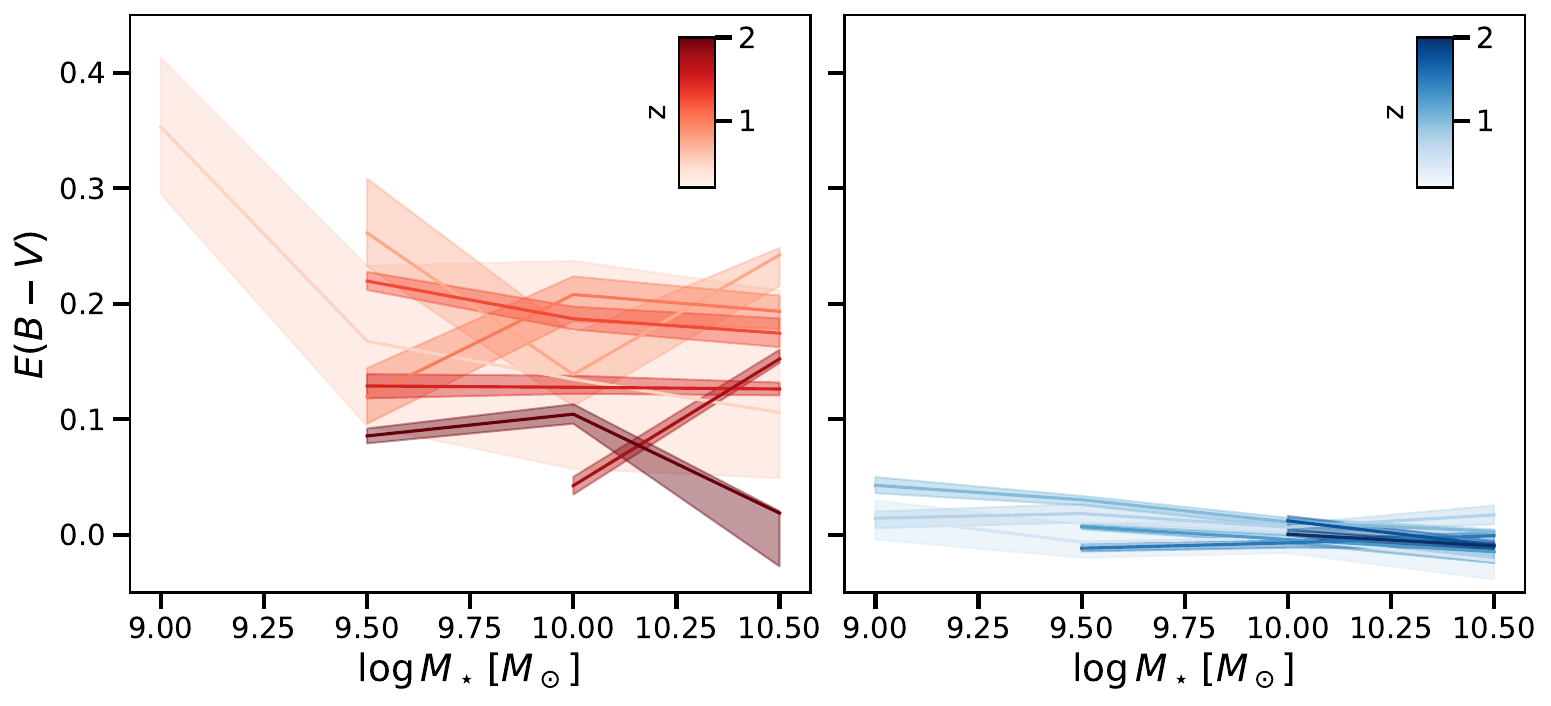}
    \caption{Same format as Figure \ref{fig:ebv_lx}. Evolution of the intrinsic optical extinction $E(B-V)$ as a function of stellar mass and redshift for rQSO (\textit{left}) and bQSO (\textit{right}). The solid lines are the median $\langle E(B-V)\rangle$ and the shaded areas show the 1$\sigma$ spread in bins of $M_\star-z$. Redshift increases with darker shades. We find no correlation between the average optical extinction and stellar mass, suggesting that the galaxy mass has little to no incidence on the dust extinction in rQSOs.}
    \label{fig:ebv_mstar}
\end{figure*}

We next investigate the host-galaxy properties to explore whether the reddening observed in quasars is linked to the stellar mass of their hosts. Stellar masses are derived through SED fitting as described in Section \ref{sec:SED}. 

Figure \ref{fig:fred_mstar} shows the fraction of rQSO, $f_{red}$, as a function of stellar mass, split into different X-ray luminosity cuts. In all luminosity bins, $f_{red}$ increases systematically with increasing stellar mass. This trend indicates that rQSO preferentially occupy massive galaxies, regardless of luminosity, which further demonstrates that the red colour is not driven solely by luminosity selection effects making it easier to detect sources in massive galaxies due to their brightness, but instead reveals an underlying dependence on host-galaxy mass.

Figure \ref{fig:ebv_mstar} shows the relation between the dust extinction, $E(B-V)$, and stellar mass of the host galaxy for rQSO on the left panel and bQSO on the right. Similar to Figure \ref{fig:ebv_lx}, the solid lines are the median $\langle E(B-V)\rangle$ and the shaded areas indicate the 1$\sigma$ spread in bins of $M_\star-z$. Redshift increases with darker shades. For both populations, we observe no dependence on stellar mass for the extinction $E(B-V)$ across the full range of masses. In the case of rQSO, the level of extinction remains approximately constant from low to high-mass hosts, despite the increase in rQSO fraction visible in Figure \ref{fig:fred_mstar}. Similarly, blue quasars show no evidence for higher extinction in more massive galaxies.

These results demonstrate that while rQSOs are more commonly associated with massive hosts, the host stellar mass itself does not regulate the amount of dust reddening. Hence, the increase in the fraction of rQSO with stellar mass cannot be attributed to a scaling of dust content with galaxy mass; instead, the absence of a correlation between $E(B-V)$ and stellar mass implies that the reddening arises from processes that are independent of the global stellar mass of the host.

\section{Discussion}
\label{sec:discussion}
\subsection{Comparison with previous studies}
\label{sec:comparison}
Previous work on rQSOs showed contradictory results on their physical nature, mainly due to the non-homogeneous definitions and selection methods used. Most studies rely on radio or infrared-selected rQSOs, usually with no direct X-ray constraints. Major analyses such as \citealt{klindt_fundamental_2019,fawcett_fundamental_2020,fawcett_fundamental_2022,fawcett_striking_2023} have demonstrated that rQSOs have higher radio detection fractions than control bQSO samples, although radio-detected rQSOs tend to be in the radio-quiet regime. These works have also shown rQSOs to be mostly associated with compact radio morphologies, indicating that powerful but non-extended outflows are likely to be responsible for their radio emission. These works conclude that while outflow properties can differ, accretion appears to operate in a similar manner across the red and blue populations. Specifically, \citealt{fawcett_fundamental_2022} conclude that rQSOs do not appear to lack a radiatively efficient accretion disc once corrected for dust extinction and that the differences in radio emission trace the interaction between AGN-driven winds and the dusty ISM, rather than differences in accretion physics. Their enhanced radio emission is not produced in the radiatively inefficient, jet-dominated quasar mode \citep{heckman_coevolution_2014}, instead rQSOs are associated with compact radio structures, inconsistent with large-scale extended jets. However, the lack of homogenous X-ray coverage prevented the authors from reaching firm conclusions as to whether their red quasar samples also produce the X-ray emission expected in the standard AGN model. 

The study of the X-ray properties of rQSOs is mostly limited to the work of \citealt{lamassa_peering_2016} and \citealt{glikman_peering_2017}. They presented a combined sample of four luminous rQSOs at low redshift ($z<1$) and found moderate absorption with column densities of $N_H \sim 10^{23} \, \rm cm^{-2}$. Section \ref{sec:obscuration} describes in more detail their result on the dust properties compared to our sample. Relevant to this work, \citealt{goulding_high-redshift_2018} and \citealt{ma_evidence_2024} have also presented X-ray studies on reddened quasars, although these studies yield contradictory results. Both works focus on relatively small samples of extremely rQSOs (ERQs), taken from \citealt{ross_extremely_2015} and \citealt{hamann_extremely_2017}, selected based on their extremely high infrared-to-optical flux ratios as well as spectroscopic properties such as high equivalent widths of emission lines, indicative of continuum suppression, and observed with targeted X-ray observations, for 11 ERQs out of $\sim$300 in the parent samples. \citealt{goulding_high-redshift_2018} did not find differences in the intrinsic X-ray emission of ERQs compared to bQSOs and that ERQs do not appear to be intrinsically fainter in X-rays for their bolometric luminosity. However, \citealt{ma_evidence_2024} found contradictory evidence. Within their sample of 50 ERQs, they found that half of their sources lacked an individual detection in X-ray, and that once stacked, their luminosity was on average lower by a factor of 10 than expected based on their infrared luminosities, concluding that ERQs are both intrinsically X-ray weak and heavily obscured. Our approach here differs, as our sample is initially selected from all-sky X-ray surveys with \textit{eROSITA} for which we obtain optical spectra with SDSS. Thus we probe a different population, representing the relatively reddest objects from within the eROSITA X-ray sample, based on observed optical colours. By construction, our sample is required to be X-ray detected. However, our X-ray selected rQSOs tend to have lower X-ray luminosities than our control bQSOs. These differences in the X-ray luminosities remain even after correcting for X-ray absorption effects for the subsample in eFEDS where we have estimates of absorption-corrected luminosities. Thus, our work suggests that, consistent with the \citealt{ma_evidence_2024} findings for ERQs, rQSOs correspond to an intrinsically X-ray weak population as supported by our measured $\alpha_{OX}$ (Section \ref{sec:aox}), on average lower in rQSOs, indicating weaker X-ray luminosities than expected based on their optical emission compared to bQSOs.

\subsection{Origin of the absorbing material}
Our work demonstrates that the fraction of rQSOs is larger in higher stellar mass hosts, indicating that rQSOs are more likely to be hosted by higher stellar mass galaxies compared to the general QSO population (Figure \ref{fig:fred_mstar}), but no correlation between dust reddening $E(B- V)$ and stellar mass is found (Figure \ref{fig:ebv_mstar}). This key result implies that massive hosts may have the conditions required for dust production, but the reddening itself may instead stem from AGN-scale structures. Disentangling host-scale and nuclear dust is essential for understanding how dust relates to accretion, feedback, and galaxy evolution \citep{hickox_obscured_2018}. Optical reddening traces dust along the line-of-sight to the accretion disc and broad-line region, whereas X-ray absorption probes gas along the line-of-sight to the X-ray corona. Comparing these two measurements provides a powerful diagnostic of whether gas and dust are coupled or come from spatially distinct structures. If the dust is nuclear, it will probe the close environment of BHs and the geometry of the obscuring torus. However, if the dust is of host origin, it becomes a tracer of large-scale processes such as feedback-driven outflows that can temporarily obscure and suppress accretion. In Sections \ref{sec:obscuration} and \ref{sec:galfrac}, we investigate whether the dust responsible for extinction in rQSOs is of nuclear or host origin.

\subsubsection{X-ray absorption as an indicator of nuclear dust structures}
\label{sec:obscuration}
To investigate whether the dust reddening of rQSOs originates from the nuclear region, we examine their line-of-sight X-ray absorption. If the reddening was primarily due to the classical dusty torus, we would expect to find systematically higher equivalent neutral hydrogen column densities, $N_H$, in rQSOs compared to blue quasars.

In this section, we further explore the connection between the reddening due to dust measured from optical spectroscopy and line-of-sight X-ray absorption measured from X-ray spectral analysis by \citealt{liu_erosita_2022} for the eFEDS subset of our sample (720 blue and 134 red QSOs). This subset spans the same X-ray luminosity - redshift space as our parent sample, but somewhat clusters around the lower luminosity end of the distribution due to deeper exposures in eFEDS.

\begin{figure}
    \centering
    \includegraphics[width=0.8\linewidth]{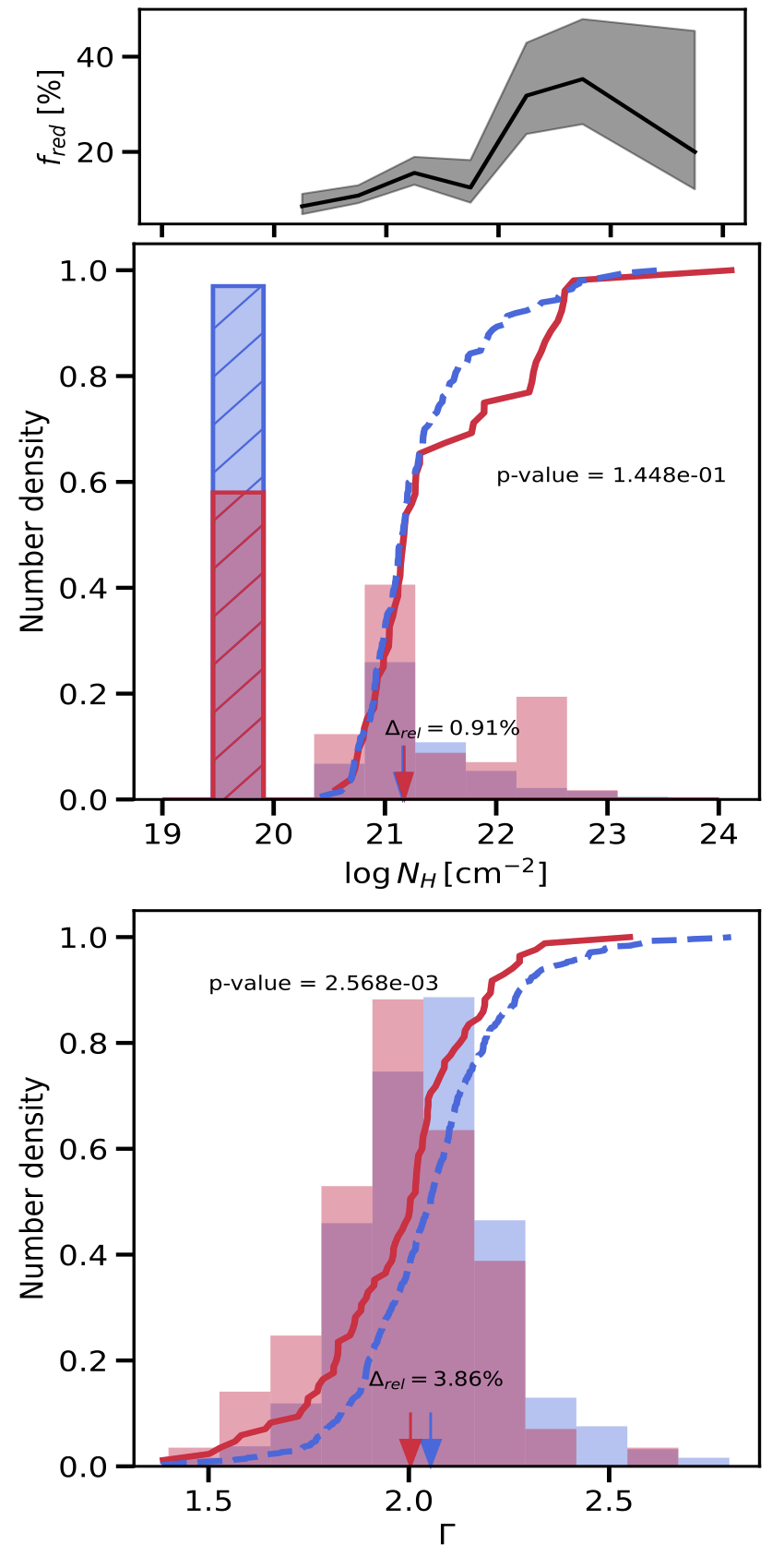}
    \caption{X-ray spectral measurements for the eFEDS sub-samples \citep{liu_erosita_2022}. \textit{Top:} Fraction of rQSOs as a function of hydrogen column density $N_H$. Below are the $N_H$ distributions for bQSOs and rQSOs. The dashed bin are ''unobscured'' sources set at $\log N_H = 19.5 \, \rm cm^{-2}$. \textit{Bottom:} Photon index $\Gamma$ distribution. The median of the distributions are shown by the arrows, and the Cumulative Distribution Functions (CDFs) are shown by the solid red and dashed blue lines for red and blue quasars respectively. We find a relative difference in $N_H$ of $\Delta_{rel} \sim 0.91 \%$ and $\Delta_{rel} \sim 3.86 \%$ in $\Gamma$. A 2-sample KS-test on the CDF produces a $p$-value of $\sim 0.1448$ and $2.568\times 10^{-3}$ respectively.}
    \label{fig:NH}
\end{figure}

The top panel of Figure \ref{fig:NH} compares the distributions of $N_H$ between the two populations using X-ray spectral measurements from \citealt{liu_erosita_2022} for the eFEDS subsample. We make use of the fitting results for their absorbed powerlaw model\footnote{Model 1.} to study the spectral properties $N_H$ and $\Gamma$. In the following, we exclude all sources with fitting results classified as ``uninformative'' by the authors, which includes $\sim10\%$ of the bQSO sample and $\sim 5\%$ of rQSOs. We retain ``unobscured'' sources where it was not possible to provide an accurate measurement of $N_H$ but the column density is constrained to be $\lesssim 10^{20}$~cm$^{-2}$. For these sources, we instead set their column density values to $N_H = 10^{19.5} \, \rm cm^{-2}$ (shown by the dashed histograms). The distributions show that both rQSOs and bQSOs reside predominantly in the Compton-thin regime, with most sources having $N_H < 10^{22}\,\rm cm^{-2}$. Both distributions peak at similar moderate values ($N_H \sim 10^{20.5-21}\,\rm cm^{-2}$), indicating that most of the rQSOs in our sample are not heavily obscured in X-ray and do not resemble classical type-2 quasars, which differs from the measurements of \citealt{ma_evidence_2024} who report much higher column density values ($N_H \sim 10^{23-24}\,\rm cm^{-2}$) based on targeted X-ray observations of an extremely red quasar sample, with higher optical reddening than our sample. As such, our results do not favour a purely geometric explanation i.e., an optically-thick dusty torus blocking our line-of-sight to the central accretion disc as the primary cause of reddening in rQSOs, that also absorbs the X-ray photons. However, we also observe an excess of rQSOs at $N_H > 10^{22}\,\rm cm^{-2}$, suggesting that for a small subset 
there is significant obscuring material along the line-of-sight to the X-ray emitting region.
This feature is absent in the blue quasar population and could reflect dust embedded within the inner AGN structure, but it does not dominate the population. 

\begin{figure}
    \centering
    \includegraphics[width=0.99\linewidth]{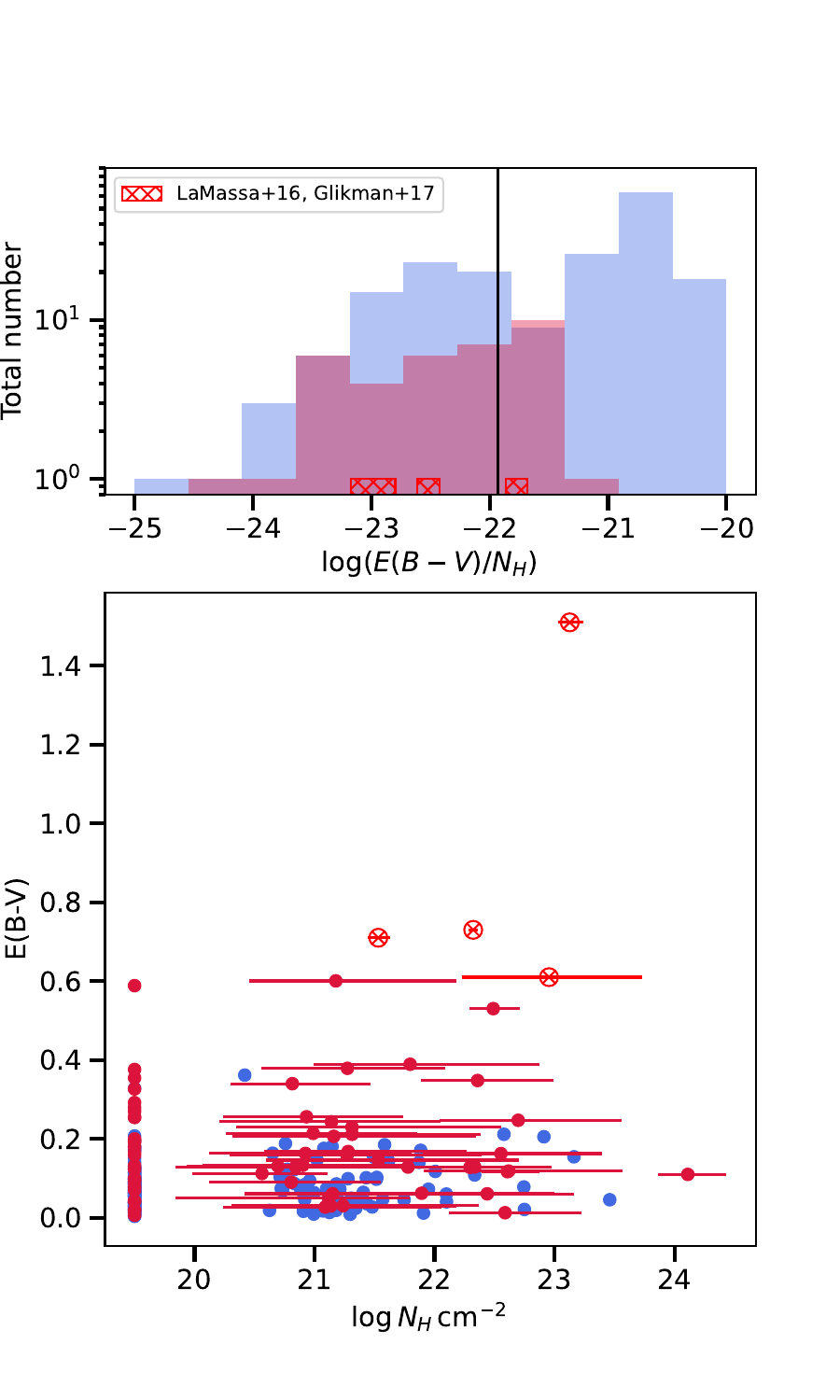}
    \caption{\textit{Top panel:} Dust-to-gas ratio distribution for rQSOs and bQSOs, compared to the rQSO samples from \citealt{lamassa_peering_2016,glikman_peering_2017}. The solid black line indicates the Milky Way dust-to-gas ratio \citep[$1.7\times 10^{-22}\, \rm cm^2$;][]{bohlin_survey_1978}. \textit{Main panel:} Relation between gas and dust absorption $E(B-V) - N_H$.}
    \label{fig:nh-ebv}
\end{figure}

Here, we only present results from \citet{liu_erosita_2022} based on their single absorbed power law model, after testing the robustness of these measurements compared to then more complex absorption models explored by \citet{liu_erosita_2022} (including a double power law i.e. a partially covering absorber model, and a power law plus blackbody emission model).
We find that $N_H$ remains consistent across models, with ratios between models roughly around 1 but with however larger deviations between the powerlaw and double power law models. It seems that more complex models lead to slightly higher column densities but this does not alter the overall distributions or the relative comparison between rQSOs and bQSOs as this effect is observed homogeneously across both populations. However, the limited photon counts in the eFEDS X-ray spectra prevents robust constraints of more complex absorption features such as ionised absorbers that cannot be captured with these models. We stress that the $N_H$ measurements we present must be interpreted as equivalent \emph{neutral} hydrogen column densities, which do not account explicitly for absorption by ionised gas along the line of sight. Thus we might underestimate the \emph{total} column density of gas, both neutral and ionised which is not captured in the definition of $N_H$. However, this effect would thus increase the true column densities of both populations, supporting that rQSOs have even larger gas reservoirs than currently measured, lowering the dust-to-gas ratio (Figure \ref{fig:nh-ebv}), further supporting that the X-ray absorption in rQSOs likely originates from dust-free gas.

The bottom panel of Figure \ref{fig:NH} displays the distribution of photon indices, $\Gamma$, for the absorbed power-law fits to the eFEDS subsamples of rQSOs and bQSOs. rQSOs show slightly lower photon indices and thus harder spectra with a median $\Gamma \sim 1.9$ versus $\Gamma \sim 2.0-2.1$ for bQSOs, consistent with the average $\Gamma$ in \textit{eROSITA} samples, indicating an intrinsically altered corona and lower Eddington ratios \citep[e.g.,][]{brightman_statistical_2013}, as previously shown in Section \ref{sec:edd}, but not consistent with typical heavily obscured type-2 AGN. Importantly, we note that the \textit{eROSITA} selection will be biased against the selection of heavily obscured sources, frequently missed due to strong absorption in the soft band where \textit{eROSITA} is the most sensitive, and thus preferentially detecting sources with soft spectra. Thus, the relatively high photon indices compared to most X-ray AGN samples \citep{nandra_xmm-newton_2007} measured for both samples could at least be partially driven by selection effects with \textit{eROSITA}. Even so, rQSOs have slightly flatter $\Gamma$ compared to bQSOs despite the selection bias toward steeper slopes that should impact both samples. This slight shift towards harder sources can arise from differences in their intrinsic coronal properties. However, the absence of strongly hard spectra ($\Gamma < 1.5$), associated with the predominantly Compton-thick $N_H$ values rules out heavy nuclear obscuration as the primary cause of reddening in our sample since rQSOs \textit{do} exist within the eROSITA-selected quasar population. Instead, our results are consistent with a scenario where rQSOs could experience a weakened accretion state, causing reduced soft X-ray emission without requiring substantial line-of-sight absorption by a dusty torus.

Figure \ref{fig:nh-ebv} shows the relation between gas absorption $N_H$ and optical reddening $E(B-V)$ for the eFEDS subsample. We find no significant correlation between X-ray absorption and optical reddening for both rQSOs and bQSOs. Moreover, rQSOs cover a wide range of $E(B-V)$ at fixed $N_H$, with sources going through substantial reddening but little to no measurable X-ray absorption. The absence of a correlation between $N_H$ and $E(B-V)$ shows that they do not trace the same absorbing medium; instead, the gas responsible for X-ray absorption and dust reddening the optical emission likely arise from different spatial structures.

These results demonstrate that the rQSOs in our sample do not have X-ray properties expected from strong nuclear absorption if the optical reddening was dominated by an optically-thick torus. The X-ray absorption likely originates from dust-free gas located close to the black hole, while the dust responsible for optical reddening is produced at greater distances. Thus, the X-ray spectral measurements rule out a scenario where the reddening in rQSOs is dominated by a classical, optically thick dusty torus. Instead, the dust originates from more scattered nuclear dusty gas structures within the AGN that could attenuate the UV/optical emission without necessarily absorbing the compact X-ray corona along most line-of-sights or from dusty structures at larger spatial scales within the host, which we explore in Section~\ref{sec:galfrac}

\subsubsection{Absorption from host-scale dust structures}
\label{sec:galfrac}

\begin{figure}
    \centering
    \includegraphics[width=1.1\linewidth]{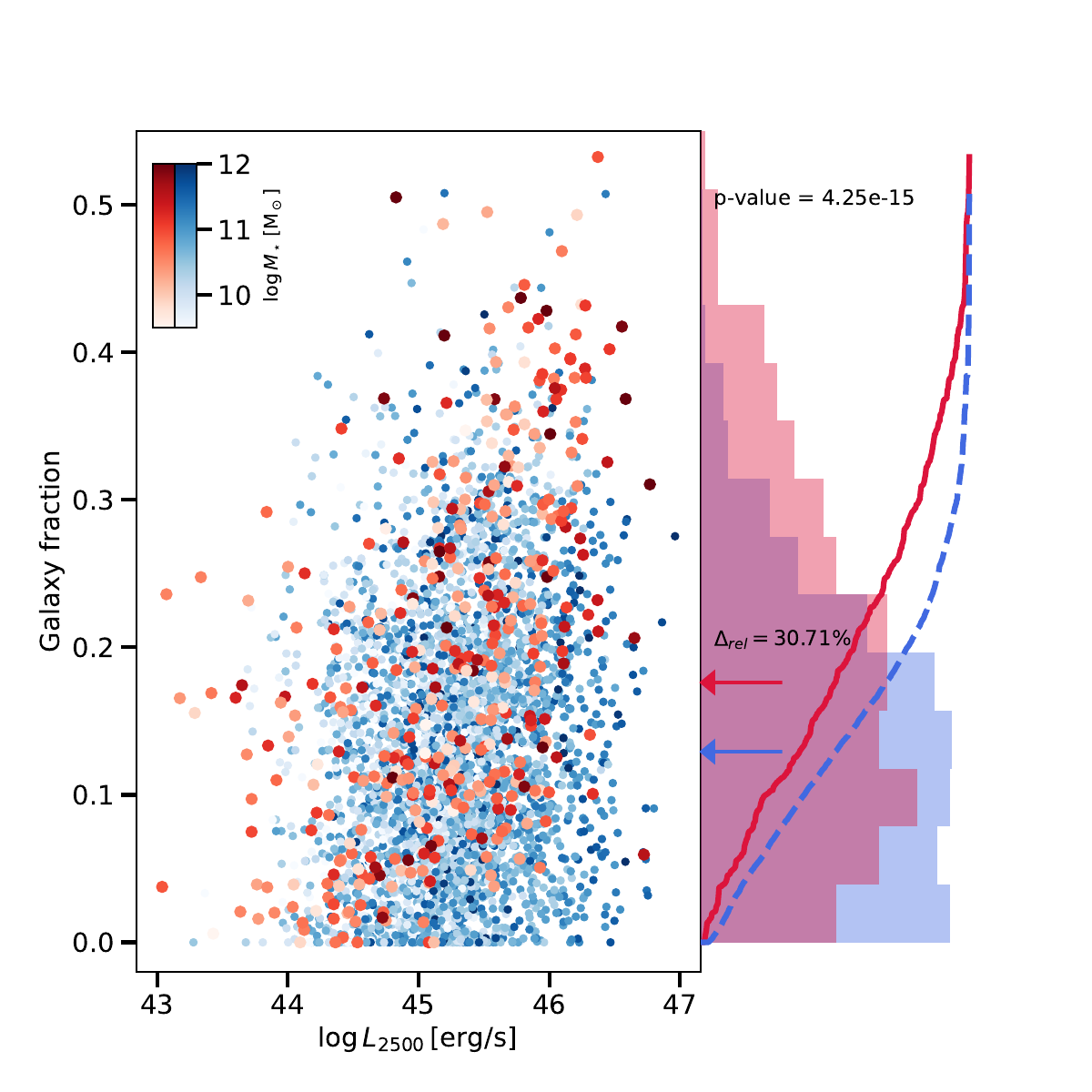}
    \caption{The total spectral galaxy fraction as a function of dust-corrected optical luminosity $L_{2500}$. Points are coloured by their stellar mass measured via SED fitting. We don't find an obvious trend between stellar mass and galaxy fraction, although the most massive galaxies tend to have higher galaxy fraction. The brighter a source is in the optical, the more host-dominated it becomes (increasing galaxy fraction with luminosity). We show on the side the distribution of the galaxy fraction for red and blue quasars. There is a striking difference between the two populations, with rQSOs showing much more often higher galaxy contributions to their optical continuum.}
    \label{fig:galfrac}
\end{figure}

To test whether dust reddening in rQSOs stems from extended host-scale structures, we use our ICA-based spectral decomposition (see Section \ref{sec:ICA}), which separates each spectrum into quasar and host-galaxy components. We quantify the host contribution as the ratio between the sum of the five galaxy component weights and the total spectral weight. The distribution of the galaxy fraction is given in Figure \ref{fig:galfrac} for red and blue quasars and its evolution with AGN optical luminosity $L_{2500}$. As expected, bQSOs are mainly AGN-dominated, with the bulk of sources having galaxy fractions typically lower than $\sim 0.2$, consistent with classical unobscured quasars where broad line emission overwhelms features from the underlying galaxy emission. In contrast, rQSOs span a significantly broader range of galaxy fraction, with a longer tail extending to galaxy fractions $\sim 0.5$, a surprisingly strong contribution for sources spectroscopically classified as QSO by the BOSS pipeline due to their broad line emission. This result indicates that, while many rQSOs are still AGN-dominated, a substantial fraction shows strong host-galaxy contributions to their optical continuum. On host-scales, dust can scatter the AGN light into the line of sight, resulting in a redder continuum. Thus, higher galaxy fractions imply a more prominent stellar or dust-scattered emission, in line with host-scale absorption or enhanced star formation. Combined with low-to-moderate absorbing column densities, $N_H$, indicating a lack of strong X-ray absorption (Section \ref{sec:obscuration}), these observations suggest that the main source of reddening in rQSOs might not be a traditional compact dusty torus within the nuclear region, but could also be associated with dust distributed on larger scales within the host, or patchy dust structures in the AGN polar regions, partially obscuring the quasar optical/UV \textit{and} X-ray emission. Importantly, the lack of strong X-ray absorption in rQSOs does not rule out a host origin scenario, as host-dust structures are likely to be more spatially extended and less optically thick than nuclear structures \citep{goulding_deep_2012,hickox_obscured_2018}. Moreover, the prevalence of rQSOs in higher stellar mass galaxies support the model where that the dusty material originates from the host itself and that the reddening cannot be attributed to a simple torus.

Taken together, the X-ray and optical measurements point to a structured absorption geometry in rQSOs, where gas and dust are not located within the same environment. The absence of correlation between $N_H$ and $E(B-V)$, combined with low to moderate column densities, suggest that the optical reddening is not dominated by a dusty torus. Instead, the X-ray absorption most likely arises from dust-free gas within the nucleus, while the dust reddening the optical light originates from larger spatial scales, either from dusty gas in the polar regions of the AGN or from the host itself, as suggested by the large galaxy fractions in rQSOs and their prevalence in massive hosts, indicating that the dusty material could be of host origin.

\subsection{Fractional black hole and stellar growth}
\label{sec:frac_growth}

Here, we first investigate the differences in terms of BH mass between red and blue quasars. Single epoch virial BH masses are measured from the \ion{Mg}{II} emission line width $FWHM_\ion{Mg}{II}$ and continuum luminosity at 2100 \AA , $L_{2100}$ reconstructed with the 10 component high-accuracy fit (see Section \ref{sec:10c}), and corrected for dust attenuation and host contribution. We adopt the scaling relation from \citealt{vestergaard_mass_2009}:

\begin{equation}
    M_{BH} = 10^{6.79} \left(\frac{FWHM_\ion{Mg}{II}}{1000 \rm km \, s^{-1}}\right)^2 \left(\frac{\lambda L_\lambda}{10^{44} \rm erg \, s^{-1}}\right)^{0.5}.
\end{equation}

\begin{figure}
    \centering
    \includegraphics[scale=0.45]{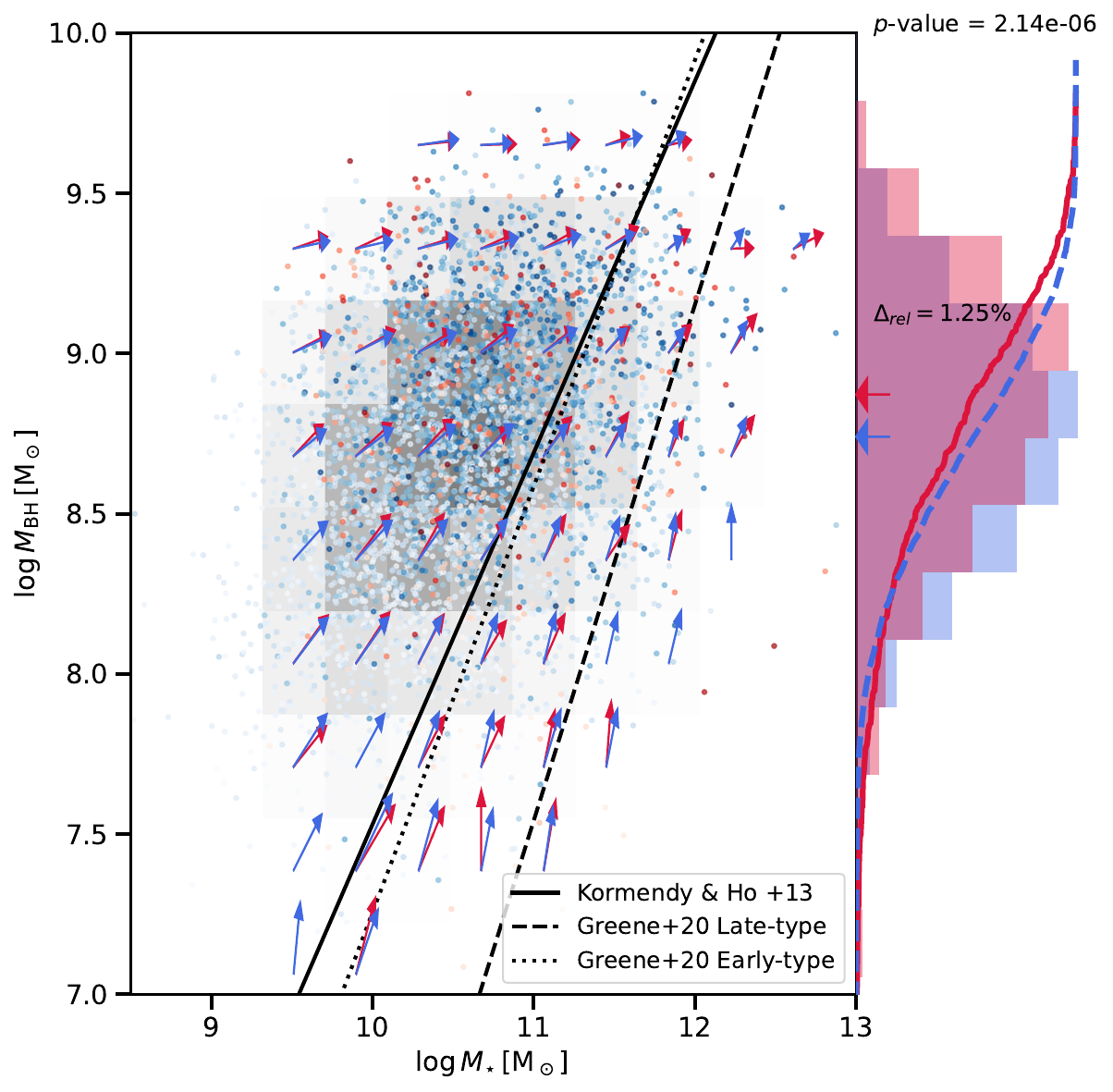}
    \caption{\textit{Main panel:} Blue and red quasars in the $M_{BH} - M_\star$ space. Points are colour-coded by redshift, a darker shade indicating higher $z$. BH masses are measured from \ion{Mg}{II} single-epoch virial masses and stellar masses are estimated via SED fitting (Section~\ref{sec:SED}). We compare our measurements to the scaling relation from \citealt{kormendy_coevolution_2013} (solid line), and \citealt{greene_intermediate-mass_2020} (early-type in dotted line and late-type in dashed line). We grid the space in bins of $M_{BH} - M_\star$ and compute for both red and blue samples the average $\lambda_{Edd}$ (y-component of the vectors) and sSFR (x-component of the vectors) measured from SED fitting. The vectors indicate the amount of BH and stellar growth in each cell and that both rQSOs and bQSOs are in general moving toward the scaling relations. \textit{Right:} Distribution of the BH mass. The median of the distributions are shown by the arrows and their relative difference $\Delta_{rel}$ is around 1.25$\%$. The Cumulative Distribution Functions (CDFs) are shown by the solid red and dashed blue lines. We perform the 2-sample KS-test on the CDFs, providing a $p$-value of 2.14$\times 10^{-6}$  consistent with the rQSO sample having, typically, higher BH masses than the bQSO sample.
    }
    \label{fig:mbh-mstar}
\end{figure}

In Figure \ref{fig:mbh-mstar}, we show the distribution of BH mass as well as the CDF in the solid and dashed lines. The medians of the distributions are indicated by the arrows. There is no striking difference in the BH mass distribution between red and blue quasars, except for a slight tail at the low end of the distribution in rQSO. However, when quantifying the significance of this difference using a 2-sample KS-test (see Section \ref{sec:LX}), the $p$-value $=2.14\times 10^{-6}$ is small enough to reject the null-hypothesis that the BH mass distributions are drawn from the same populations, but not as convincingly as $L_X$. The median of both samples are very similar with $\langle \log M_{BH}\rangle_{red} \sim 8.84 \, \rm M_\odot$ and $\langle \log M_{BH}\rangle_{blue} \sim 8.73\, \rm M_\odot$, and well within the intrinsic scatter in \ion{Mg}{II}-based BH single-epoch BH masses of $\sim 0.3 - 0.4$ dex. Thus, the statistically significant KS-test result could be primarily driven by the large sample size rather than a physically meaningful mass difference. The medians have a relative difference $\Delta_{rel}$ of about 1.25$\%$.

The main panel of Figure \ref{fig:mbh-mstar} illustrates how BHs and their host galaxies are currently growing across the $M_{BH} - M_\star$ plane for red and blue quasars. Due to our selection criteria requiring a relatively bright AGN component, and thus selecting preferentially higher BH mass quasars, we tend to find both rQSOs and bQSOs above the scaling relations \citep[black lines:][]{kormendy_coevolution_2013,greene_intermediate-mass_2020}, i.e., our sample misses low-mass BHs in massive galaxies. To understand this bias, we create a 2D grid shown by the shaded grey areas. The vectors represent the average fractional growth rates in each $M_{BH} - M_\star$ bin, where the horizontal and vertical components correspond to the average specific star-formation rate sSFR measured via SED fitting (normalized by stellar mass) and Eddington ratio $s\lambda_{Edd}$ measured from dust-corrected optical luminosities $L_{3000}$ (Section \ref{sec:edd}), respectively. As noted in Section \ref{sec:SED}, AGN contribution to the SED can lead to uncertainties in SFR measurements. In Appendix \ref{app:A}, we compare our growth vectors based on the measured SFRs with alternative estimates using prior measurements of the main sequence of star formation, finding consistent trends that indicate --- at least on average within a $M_\star-M_{\rm BH}$ bin --- that our estimates are reasonable. Each vector thus traces the instantaneous growth and evolution of sources in this space, i.e., the relative rates at which BHs and galaxies are assembling their mass, as a function of stellar and BH mass. The length of the arrow corresponds to the rate at which the galaxies in a given bin are growing both their stellar and BH components, relative to their current stellar and BH mass. Individual sources are coloured by their spectroscopic redshift, with darker shades indicating higher redshift.

Both populations show vectors that broadly converge towards the main $M_{BH} - M_\star$ scaling relations. \citealt{sun_evolution_2015} and \citealt{zhuang_evolutionary_2023} have shown similar evolutionary pathways with a persistent motion, suggesting a self-regulated mode of co-evolution where BH and stellar growth happen in tandem, such that, on average, sources evolve towards the local scaling relation regardless of their offset from it.  Indeed, for sources below the scaling relations (under-massive BH or over-massive galaxy), their evolution is mostly driven by BH growth, rising vertically to reach the relation. The opposite occurs for sources above the line, now with over-massive BHs ceasing significant BH mass assembly, while the galaxy appears to continue building stellar mass to reach the relation. 

\begin{figure}
    \centering
    \includegraphics[width=0.95\linewidth]{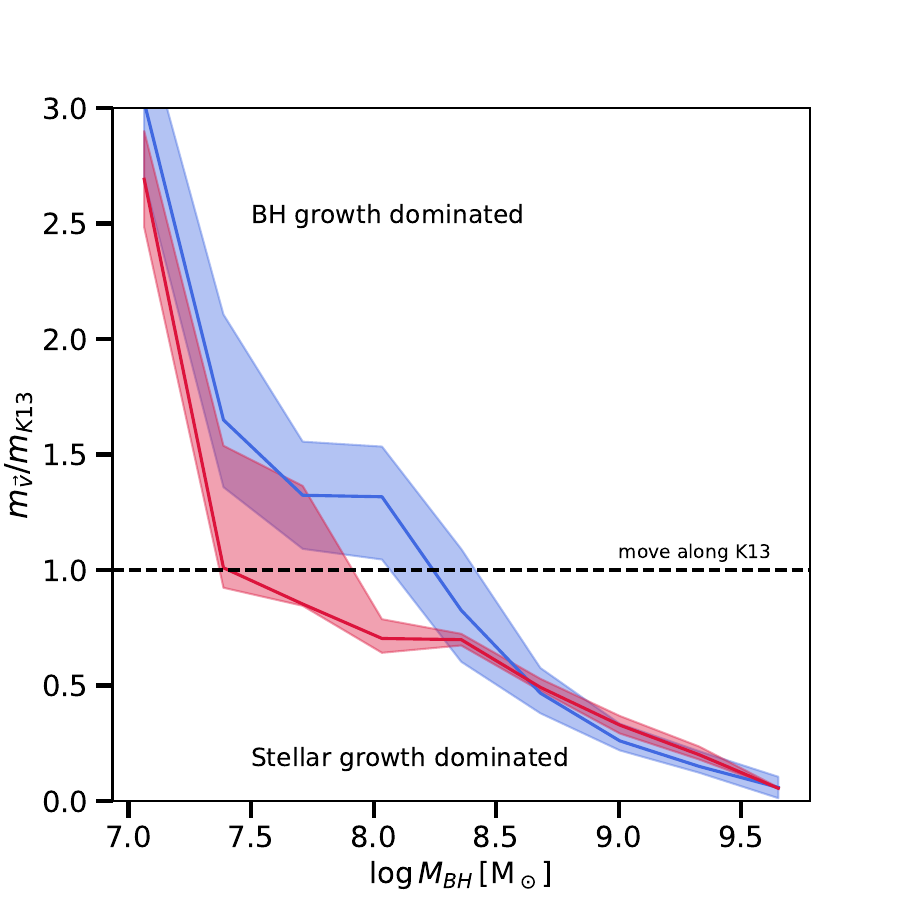}
    \caption{The slope ratio between $m_{K13}$ from \citet[][K13]{kormendy_coevolution_2013} and $m_{\vec{v}}$, the average growth vectors across stellar mass from Figure \ref{fig:mbh-mstar}, against BH mass. A ratio of one indicates an evolution along K13, above one indicates a steeper vector and thus more fractional BH growth compared to K13, whereas below one vectors are shallower and dominated by stellar growth.}
    \label{fig:slope}
\end{figure}

However, the individual behaviour differs between the two quasar populations. For bQSOs, arrows are typically longer and more vertical, indicating a higher fractional BH growth relative to their hosts. bQSOs are accreting efficiently (higher Eddington ratios) and are capable of rapidly increasing their BH mass compared to their stellar mass. Their evolutionary vectors often lie parallel to or slightly steeper than the \citealt{kormendy_coevolution_2013} relation, implying that bQSOs can maintain or even increase their BH-to-stellar-mass ratio if their current growth is sustained. In contrast, rQSOs show systematically shorter arrows with lower slopes. Their vertical components, which trace normalized BH accretion, are smaller, consistent with their on-average lower Eddington ratios and lower X-ray luminosities found in Section \ref{sec:LX}. This behaviour indicates that rQSOs are accreting less efficiently, despite similar stellar growth rates. Thus, their evolutionary vectors are more horizontal, pointing towards a phase in which the host galaxy grows faster than the central BH. 

Figure \ref{fig:slope} quantifies these trends by comparing the slope of the growth vectors ($m_{\vec{v}} = \Delta\log M_{BH} / \Delta\log M_\star = \lambda_{Edd}/sSFR$) to the slope of the \citealt{kormendy_coevolution_2013} relation $m_{K13}$. The slope ratio measures how the instantaneous direction of growth compares to the one required to remain on the scaling relation. A ratio of one (dashed line) corresponds to growth parallel to the relation, preserving the same growth rate predicted by the empirical relation. Ratios greater than one indicate relatively stronger BH growth, i.e., sources becoming over-massive if the ratio is sustained, while values lower than one mean stronger stellar mass increase. 

Taking the average slope across stellar mass for each BH mass bin, Figure \ref{fig:slope} shows the evolution of the fractional BH growth rates with BH mass. Both populations show a clear decrease of the slope ratio with increasing BH mass. At low BH masses $(\log (M_{BH}/\rm M_\odot) \leqslant 7.5 - 8)$, both red and blue QSOs have ratios above one, implying that low-mass BHs are currently in a phase where BHs grow faster relative to their hosts, potentially catching up to the scaling relation. At higher BH masses, the ratio drops below one, and more rapidly for rQSOs. This result indicates that once the BH reaches $\sim \, \rm 10^{7.5-8} M_\odot$, its fractional accretion rate declines relative to the sSFR of the host, a sign of a transition to a BH-regulated growth mode. In this regime, galaxy growth proceeds faster than BH accretion, leading to an evolutionary track that moves sources towards the scaling relation. This finding is consistent with the need for rapid early BH assembly to reach the local relation by $z\sim 0$ to produce massive BHs above the scaling relations, so they do not have to grow their mass as much at later times \citep{guetzoyan_x-ray_2024,terrazas_diverse_2025}. At fixed BH mass below $10^{8.5}\,\text{M}_\odot$, bQSOs show systematically higher fractional BH growth than rQSOs, reflecting the different phases of BH fuelling within the lifetime of a quasar. Indeed, tracking BH assembly in time shows that quasars undergo a sharp rise in accretion rate in the early phases of growth to reach the most efficient accretion regime where most of their mass is assembled. This early rise corresponds to the rQSO phase, (having systematically lower slope ratios and thus less fractional BH growth than bQSOs at fixed BH mass), which is succeeded by a more sustained and efficient accretion state, when transitioning to the bQSO phase. This model suggests that bQSOs are observed during the peak of BH growth activity before slowing down their BH mass assembly once high BH masses are reached, but continuing to grow their stellar mass, while rQSOs could be consistent with capturing an earlier dust-enshrouded stage of BH growth, when the absolute BH growth is lower \citep{hopkins_unified_2006,alexander_what_2025}. 

Moreover, the more pronounced and earlier drop with BH mass for rQSOs suggests that they transition to a host-dominated build-up phase at lower BH mass than bQSOs. This proposal suggests that for low-mass BHs to continue building mass and eventually reach the expected local scaling relation, they need to transition from a red to a blue quasar phase, where BH growth becomes more efficient and catches up to the assembly of the host. This interpretation is consistent with the broader evolutionary scenario of rQSOs, explained as transitional phases between the obscured, dust-rich growth phase and the unobscured luminous QSO phase. The rQSOs' suppressed accretion rates and flattened evolutionary vectors suggest that they are transitioning to a more active phase of BH growth.

Taken together, Figures \ref{fig:mbh-mstar} and \ref{fig:slope} support a scenario where red and blue quasars trace distinct phases of BH–galaxy assembly. bQSOs most often represent a growth-dominated phase where BH accretion overtakes stellar mass increase, while rQSOs correspond to a less evolved or transitional stage in which the galaxy continues forming stars but BH growth is suppressed, potentially by strong feedback processes during a blow-out phase.

\subsection{Implications for an evolutionary scenario}

\begin{figure}
    \centering
    \includegraphics[width=\linewidth]{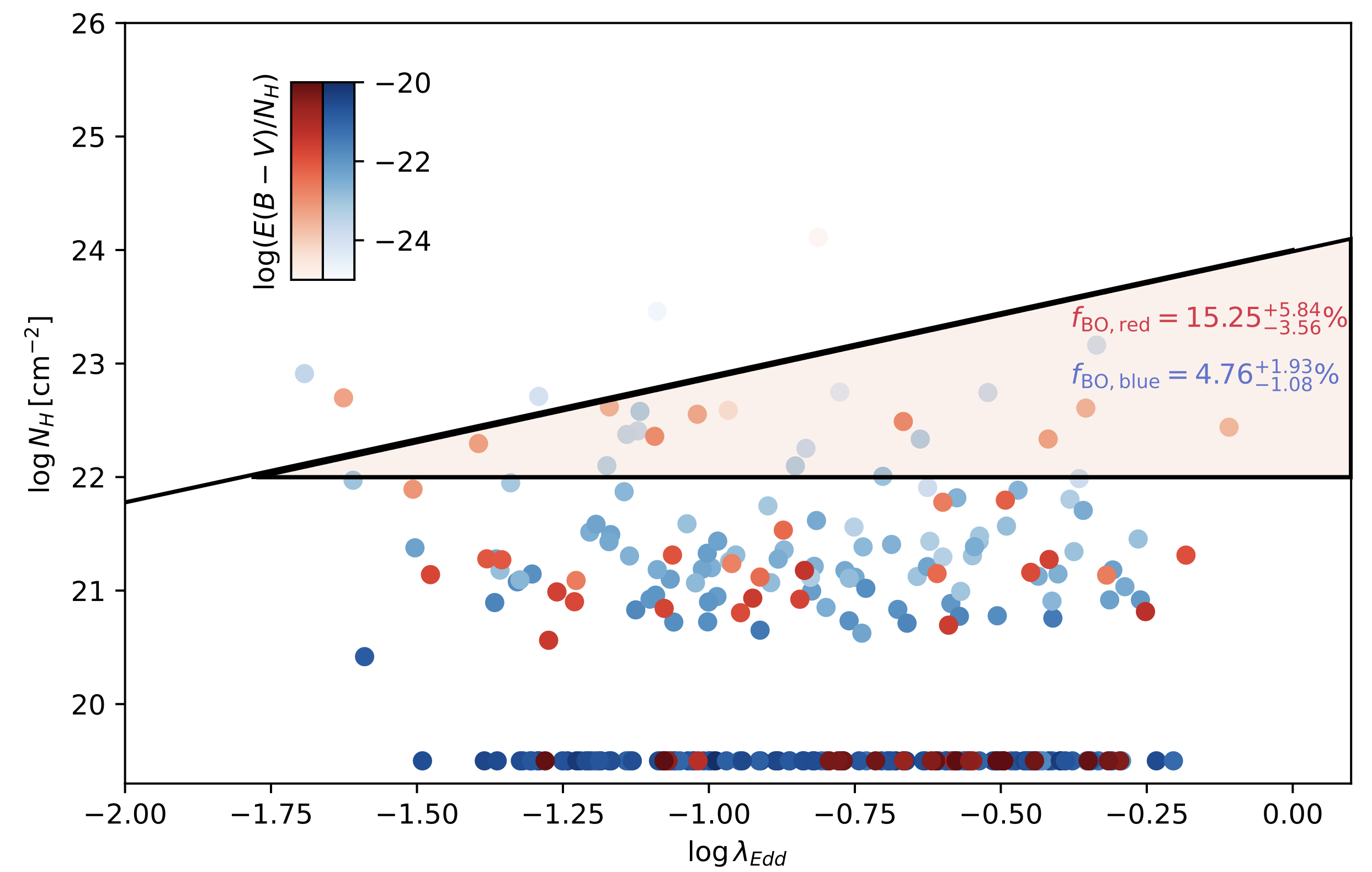}
    \caption{Column density $N_H$ - Eddington ratio $\lambda_{\rm Edd}$ diagram. The horizontal black line is the $N_H = 10^{22}\, \rm cm^{-2}$ limit, while the intersecting black line represents the effective Eddington limit for dusty gas. The area in-between corresponds to sources with $\lambda_{\rm Edd}$ higher than the effective limit for a given $N_H$, i.e., able to expel the surrounding material due to higher radiation pressure. This region is denoted as the ``blow-out'' phase \citep{ricci_close_2017} and shown by the light orange area. A higher fraction of sources in the blow-out region are from our rQSO sample. Points are coloured by their dust-to-gas ratio value.}
    \label{fig:blowout}
\end{figure}

Figure \ref{fig:blowout} shows the $N_H - \lambda_{\rm Edd}$ diagram, which has been proposed as a means of identifying AGN at different stages in a feedback cycle regulating obscuration, with sources potentially transitioning from obscured to unobscured phases. The solid black lines (adapted from \citealt{ricci_close_2017}) show where host-scale dust is expected to contribute to the column density versus torus obscuration above the horizontal limit (fixed limit at $N_H = 10^{22} \, \rm cm^{-2}$), and the effective Eddington limit ($\lambda^{\rm eff} _{\rm Edd}$) for dusty gas. The region in-between is defined as the \textit{blow-out} region. Indeed, the cross-section of material made up of gas and dust is higher than that of pure neutral gas alone, as assumed when measuring $N_H$. Thus, radiation pressure applied on dusty gas is much stronger than predicted by the classical Eddington limit in the absence of dust. Sources with $\lambda_{\rm Edd} > \lambda^{\rm eff} _{\rm Edd}$ thus produce sufficient radiation pressure to expel obscuring material, placing them in the blow-out region.

Relative to the total number of rQSOs and bQSOs, rQSOs preferentially populate this blow-out region, with a fraction of rQSOs in the blow-out region, $f_{\rm BO,red}$ of more than $15\%$, against $f_{\rm BO,blue} \sim  5\%$ of bQSOs. We test the significance of this difference by computing the confidence intervals on fractions from \citealt{cameron_estimation_2011} and find that the ranges of blow-out fraction for both populations do not overlap and are statistically significant at the $1\sigma$ level. In absolute terms, 22 quasars fall within this space, including 9 and 13 red and blue quasars respectively, corresponding to roughly $8\%$ of the total quasar sample. Although, rQSOs are less numerous by definition, they occupy the blow-out region nearly as much as bQSOs and constitute over double the fraction of their own population within this phase. The presence of sources in this region suggests they have a larger dust content than predicted by the column density measurements, enabling radiation pressure to reach high enough values to drive outflows and expel surrounding obscuring material. 

There is also an inhomogeneous distribution of sources in the $N_H - \lambda_{\rm Edd}$ space, with the bulk of sources clustering below the $N_H = 10^{22} \, \rm cm^{-2}$ limit, while a smaller subset appears directly in the blow-out region with no clear continuum between the two regimes. This clustered distribution of sources supports that the blow-out phase is relatively short-lived, with sources rapidly evolving in and out of this stage.

\begin{table*}
    \centering
    \begin{tabular}{c|c|c|c|c|c}
         & $\langle$Red$\rangle$ & $\langle$Blue$\rangle$ & $p$-value & $\Delta_{rel} \, (\%)$ & \\
         \hline
         $L_X \, \rm [erg \, s^{-1}]$ & 4.8$\times 10^{44}$  & 5.3$\times 10^{44}$ & 2.698$\times 10^{-2}$  &  8.98 & Fig~\ref{fig:LX} \\
         \hline
         $\log \lambda_{Edd}$ & -0.91 & -0.81 & 3.12$\times 10^{-7}$ & 11.46 & Fig~\ref{fig:edddistrib} \\
         \hline
         $L_{2500}\, \rm [erg \, s^{-1} \, Hz^{-1}]$ & 1.78$\times 10^{30}$ & 1.49$\times 10^{30}$ & 4.972$\times 10^{-5}$ & 16.83 & Fig~\ref{fig:aox} \\
         \hline
         $\alpha_{ox}$ & -1.45 & -1.41 & 6.68$\times 10^{-7}$ & 3.25 & Fig~\ref{fig:aox} \\
         \hline
         $\Delta \alpha_{ox}$ & 0.064 & 0.035 & 3.79$\times 10^{-5}$ & 57.58 & Fig~\ref{fig:aox} \\
         \hline
         $M_\star \, \rm [M_\odot]$ & $10^{10.70} $& $10^{10.53}$ & 1.816$\times 10^{-7}$ & 1.57 &  Fig~\ref{fig:fred_mstar}\\
         \hline
         $\log (N_H/\rm cm^{-2})$ & 20.9 & 20.7 & 1.448$\times 10^{-1}$ & 0.91 & Fig~\ref{fig:NH} \\
         \hline
         $\Gamma$ & 2.0 & 2.05 & 2.568$\times 10^{-3}$ & 3.86 & Fig~\ref{fig:NH} \\
         \hline
         $M_{BH} \, \rm [M_\odot]$ & $10^{8.84} $ & $10^{8.73} $ & 2.14$\times 10^{-6}$ & 1.25 & Fig~\ref{fig:mbh-mstar} \\
         \hline

    \end{tabular}
    \caption{Comparison of medians between red and blue quasars (first two columns), $p$-values and relative differences $\Delta_{\rm rel}$ for all parameters explored in this work.}
    \label{tab:placeholder}
\end{table*}

Our work supports the model that rQSOs differ fundamentally from typical bQSOs in their accretion properties. rQSOs show lower intrinsic X-ray luminosities for a given optical luminosity (see Section \ref{sec:aox}) and have relatively harder photon indices $\Gamma$, while, for our selection at least, having relatively low levels of X-ray absorption ($N_H\lesssim10^{22} \, \rm cm^{-2}$) (see Section \ref{sec:obscuration}). These are signatures of suppressed corona emission rather than orientation effects due to gas absorption, indicating a lower accretion state. Recent radio analysis demonstrated that rQSOs launch strong compact outflows and exhibit enhanced radio emission compared to bQSOs, evidence for dusty winds creating shocks as it impacts the ISM in an early AGN ``blow-out'' phase \citep{fawcett_fundamental_2022,fawcett_striking_2023}. These results support a scenario where rQSOs represent a distinct phase in the lifecycle of AGN when feedback has cleared out the surrounding dust and gas from the central region. The lower X-ray luminosities and Eddington ratios in rQSOs strengthen this evolutionary scenario. Indeed, rQSOs reside in a systematically X-ray weaker regime, implying a lower accretion mode occurring in different stages of BH growth. Moreover, the larger host-galaxy contributions in rQSOs are consistent with the dust that is responsible for the reddening of the optical quasar light being distributed throughout the host galaxy rather than confined to the torus. This result suggests that feedback-driven winds could temporarily attenuate the accretion disc and suppress coronal emission, regulating BH growth. Our results support an evolutionary sequence where rQSOs are transitioning from an obscured, feedback-dominated phase toward unobscured luminous bQSOs. Thus, they represent a key stage in the self-regulated growth cycle of BHs. This evolution model also suggests a continuum with recently discovered Little Red Dots \citep[LRDs,][]{labbe_population_2023,maiolino_small_2024,greene_uncover_2024} at high redshift, where their common compact morphology, extreme reddening, and X-ray weakness may trace early-universe analogues of this same phase.

\section{Summary and Conclusions}
\label{sec:conclusions}
In this work, we have presented the first large multi-wavelength study of X-ray selected rQSOs compared to a robust control sample. Our main conclusions are as follows:

\begin{itemize}

    \item We compiled the largest X-ray sample of rQSOs to date, based on their observed $g-r$ colour excess, relative to a parent sample of $\sim$4,000 spectroscopically-classified QSOs. Taking the 10$\%$ reddest sources in bins of redshift as the rQSO sample, we selected in total 380 rQSOs and 3763 bQSOs based on their X-ray emission and optical spectroscopic observations.

    \item We successfully reconstructed quasar spectra while accounting for heavy dust reddening in rQSOs and host contribution using a novel ICA-based data-driven approach, with over 90$\%$ accuracy. With our non-parametric approach, we recovered the true distribution of optical reddening $E(B-V)$ in rQSOs to measure intrinsic optical luminosities, BH masses, and Eddington ratios. Our photometric selection of rQSOs is consistent with our measured $E(B-V)$ values, finding a median of 0.21 in rQSOs and 0.04 in bQSOs, with very little overlap between the two distributions, enabling a robust separation of the two populations using spectroscopic selection.

    \item We found an over-density of rQSOs at faint X-ray luminosities ($L_X \lesssim 10^{44}\, \rm erg \, s^{-1}$) compared to bQSOs. Furthermore, we measure a high rQSO fraction at low X-ray luminosities that decreases as $L_X$ increases, regardless of absorption correction, indicating that rQSOs are more prevalent at low $L_X$ and are thus in an intrinsically weak accretion state.

    \item The most statistically significant difference between rQSOs and bQSOs is in terms of Eddington ratios based on $L_{3000}$ continuum luminosities and \ion{Mg}{II} BH masses, with a relative difference of over 11$\%$ between the median of the two distributions. The relative number of rQSOs is higher than bQSOs at low Eddington ratios ($\lambda_{\rm Edd}\lesssim 0.1$).

    \item We investigated the X-ray weakness in rQSOs we found previously, relative to their optical emission by measuring their $\alpha_{OX}$ compared to bQSOs. There is a shift towards lower $\alpha_{OX}$ values for rQSOs, indicating a weaker X-ray emission than expected based on their optical luminosity $L_{2500}$. rQSOs do not tend to have lower ($L_{2500}$) values compared to bQSOs, which confirms that rQSOs are not just faint sources but have specifically attenuated X-ray emission due to a weaker accretion regime.

    \item rQSOs hosts show significant differences to bQSOs. A higher fraction of rQSOs reside in the most massive galaxies ($M_\star \geq 10^{10.5}\, \rm M_\odot$) and have a larger contribution to their spectral continuum, but no correlation is found between stellar mass and optical reddening $E(B-V)$.

    \item Based on X-ray spectral measurements from the eFEDS survey for a subset of our sources, rQSOs have lower dust-to-gas ratio than the Milky Way, indicating a greater gas column density at fixed extinction, absorbing X-ray emission along the line-of-sight. Moreover, there is no correlation between X-ray absorption by gas $N_H$ and optical reddening by dust $E(B-V)$, indicating that the material responsible for X-ray and optical absorption might not stem from the same structures and spatial scales.

    \item The high gas column densities in rQSOs indicate that X-ray absorption is dominated by dust-free gas located close to the nucleus, rather than the same dusty structures responsible for the optical reddening.

    \item Alternatively, the absorbing material responsible for optical reddening could be distributed on larger scales, either from dusty gas structures distributed further from the nucleus carried by disc winds or within the host galaxy itself. This scenario is supported by the higher prevalence of rQSOs in massive galaxies, and the enhanced galaxy fraction in the continuum of rQSOs spectra.

    \item Combining Eddington ratios and SFRs, we found that rQSOs are in a growth regime where BH assembly is suppressed relative to stellar mass growth. Compared to bQSOs, rQSOs are more often associated with a stellar growth-dominated phase, suggesting they represent a distinct evolutionary stage where efficient BH accretion is temporarily suppressed while the host galaxy continues to grow.
    
\end{itemize}

This work presents the first study of a large and homogeneous X-ray selected rQSO sample from the optical spectroscopy targetting program of \textit{eROSITA} sources with SDSS-V. We robustly identified $\sim$ 400 rQSOs based on a redshift-dependent photometric selection. Compared to a control sample of bQSOs, we found that rQSOs are intrinsically X-ray weaker relative to their dust-corrected optical emission. We placed indirect constraints on dust properties based on X-ray and optical absorption and found that the obscuring material responsible for optical reddening in rQSOs is likely decoupled from the material absorbing X-ray emission and must arise from different regions. We propose that the mild X-ray absorption in rQSOs stems from dust-free gas close to the BH, while optical reddening could be caused by dusty gas transported outwards in disc winds. However, high-resolution imaging and mid-IR spectroscopy would be necessary to properly disentangle the contribution and origin of the different obscuring material.

\section*{Acknowledgements}

PG acknowledges funding from an STFC studentship. JA acknowledges support from a UKRI Future Leaders Fellowship (grant code: MR/Y019539/1). ALR acknowledges support from a Leverhulme Early Career Fellowship.

Funding for the Sloan Digital Sky Survey V has been provided by the Alfred P. Sloan Foundation, the Heising-Simons Foundation, the National Science Foundation, and the Participating Institutions. SDSS acknowledges support and resources from the Center for High-Performance Computing at the University of Utah. SDSS telescopes are located at Apache Point Observatory, funded by the Astrophysical Research Consortium and operated by New Mexico State University, and at Las Campanas Observatory, operated by the Carnegie Institution for Science. The SDSS web site is \url{www.sdss.org}.

SDSS is managed by the Astrophysical Research Consortium for the Participating Institutions of the SDSS Collaboration, including the Carnegie Institution for Science, Chilean National Time Allocation Committee (CNTAC) ratified researchers, Caltech, the Gotham Participation Group, Harvard University, Heidelberg University, The Flatiron Institute, The Johns Hopkins University, L'Ecole polytechnique f\'{e}d\'{e}rale de Lausanne (EPFL), Leibniz-Institut f\"{u}r Astrophysik Potsdam (AIP), Max-Planck-Institut f\"{u}r Astronomie (MPIA Heidelberg), Max-Planck-Institut f\"{u}r Extraterrestrische Physik (MPE), Nanjing University, National Astronomical Observatories of China (NAOC), New Mexico State University, The Ohio State University, Pennsylvania State University, Smithsonian Astrophysical Observatory, Space Telescope Science Institute (STScI), the Stellar Astrophysics Participation Group, Universidad Nacional Aut\'{o}noma de M\'{e}xico, University of Arizona, University of Colorado Boulder, University of Illinois at Urbana-Champaign, University of Toronto, University of Utah, University of Virginia, Yale University, and Yunnan University.

This work is based on data from eROSITA, the soft X-ray instrument aboard SRG, a joint Russian-German science mission supported by the Russian Space Agency (Roskosmos), in the interests of the Russian Academy of Sciences represented by its Space Research Institute (IKI), and the Deutsches Zentrum f\"{u}r Luft- und Raumfahrt (DLR). The SRG spacecraft was built by Lavochkin Association (NPOL) and its subcontractors, and is operated by NPOL with support from the Max Planck Institute for Extraterrestrial Physics (MPE). The development and construction of the eROSITA X-ray instrument was led by MPE, with contributions from the Dr. Karl Remeis Observatory Bamberg \& ECAP (FAU Erlangen-Nuernberg), the University of Hamburg Observatory, the Leibniz Institute for Astrophysics Potsdam (AIP), and the Institute for Astronomy and Astrophysics of the University of T\"{u}bingen, with the support of DLR and the Max Planck Society. The Argelander Institute for Astronomy of the University of Bonn and the Ludwig Maximilians Universit\"{a}t Munich also participated in the science preparation for eROSITA.

\section*{Data Availability}
All datasets used in this work have been made public and are part of SDSS-V DR19 \citep{sdss_collaboration_nineteenth_2025} which includes the DL1 catalogue containing level one X-ray measurements from eRASS1 \citep{merloni_srgerosita_2024}. X-ray spectral measurements for eFEDS sources were taken from \citealt{liu_erosita_2022}. The measurements derived in this work are available from the corresponding author by request.



\bibliographystyle{mnras}
\bibliography{references,ref} 




\appendix

\section{Empirical Main Sequence Star Formation Rate}
\label{app:A}

\begin{figure}
    \centering
    \includegraphics[width=\linewidth]{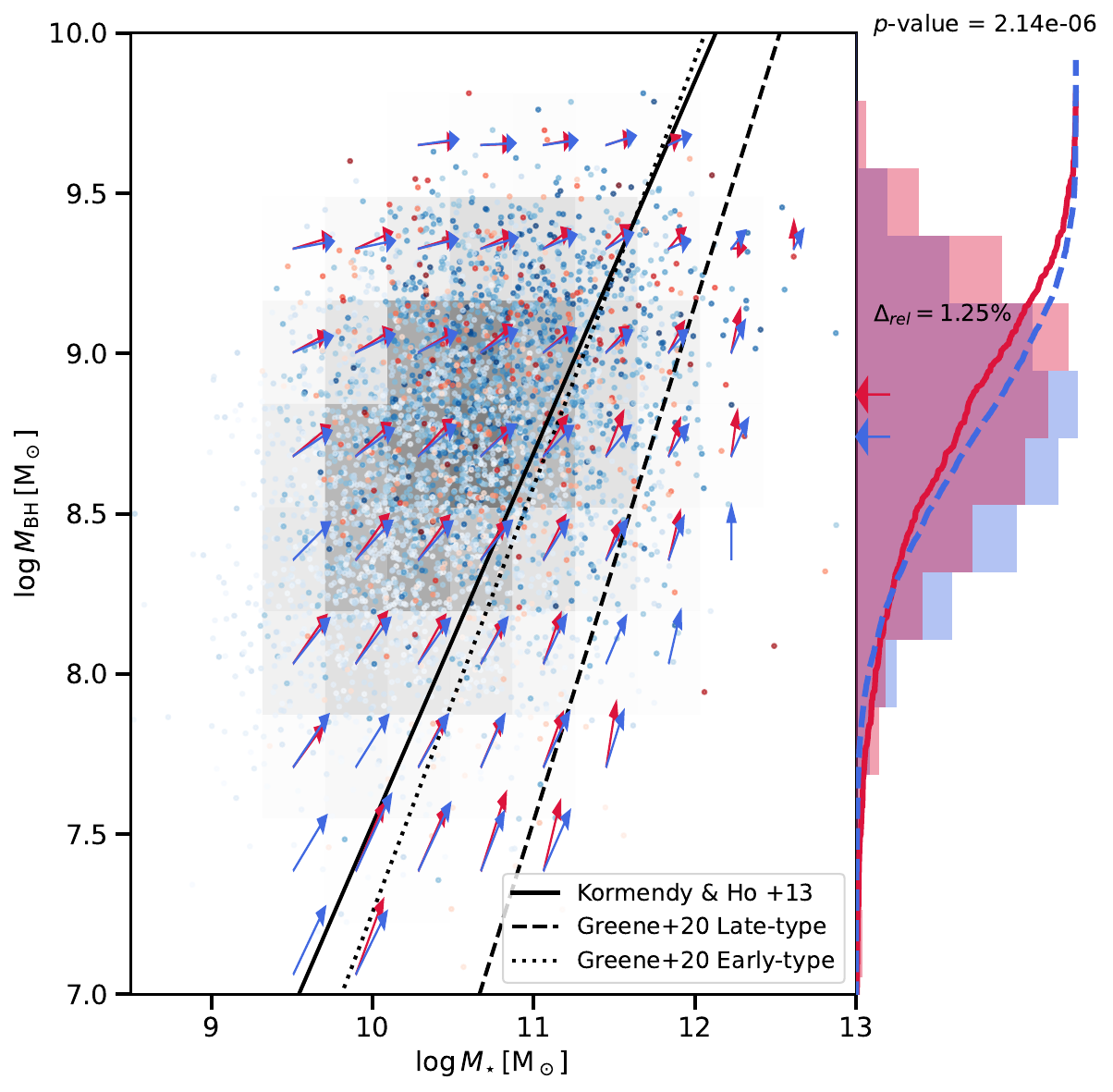}
    \caption{Same as Figure \ref{fig:mbh-mstar}. The x$-$component of the growth vectors is replaced with the average main sequence SFR from \citet[][]{popesso_main_2022} as a function of $M_\star - z$.}
    \label{fig:MS}
\end{figure}

In Sections \ref{sec:SED} and \ref{sec:frac_growth}, we discuss the reliability of our measured SFR from \textsc{cigale} SED fitting, particularly the possible bias introduced by AGN emission. Figure \ref{fig:MS} reproduces the analysis shown in Figure \ref{fig:mbh-mstar}, now using Main Sequence (MS) SFR values from \citealt{popesso_main_2022} instead of our SED-inferred SFR measurements. Overall, our main conclusions hold, regardless of the method used to estimate SFR. Assuming a MS star-formation does not significantly alter our results on BH-galaxy assembly discussed previously, we still find that rQSOs and bQSOs evolve across the $M_{\rm BH} - M_\star$ space by converging towards the scaling relations, and transitioning from BH to stellar-growth dominated phases depending on their offset from local scaling relations. This supports that average SFR measurements via SED fitting our reliable enough to be considered in this work.

\section{SED fitting parameters}

In this Appendix, we share the full parameter grid used in the SED fitting with \textsc{cigale} for both rQSO and bQSO samples (see Table \ref{cigale_param}).

\begin{table*}
	\centering
	\caption{\textsc{cigale} parameter values used in each module. This grid leads to $\sim$ 1.5 and 1.1 billion models for the bQSO and rQSO samples respectively.}
	\label{cigale_param}
	\begin{tabular}{lcr} 
		\hline
		Module & Parameter & Value\\
		\hline
		\multirow{6}{16em}{\ascii{sfhdelayed}} & \ascii{tau\_main} ($\rm Myr$) & 1,50,200,800,1000\\
                           & \ascii{age\_main} ($\rm Myr$) & 100, 200,1000, 2000, 3000, 4000\\
                           & \ascii{tau\_burst} ($\rm Myr$) & 50\\
                           & \ascii{age\_burst} ($\rm Myr$) & 20\\
                           & \ascii{f\_burst} & 0\\
                           & \ascii{sfr\_A} & 1\\
        \hline
		\multirow{3}{16em}{\ascii{bc03}} & \ascii{imf} & 1 (Chabrier)\\
                                        & \ascii{metallicity} & 0.02\\
                                        & \ascii{separation\_age} ($\rm Myr$) & 10\\
        \hline
		\multirow{6}{16em}{\ascii{nebular}} & \ascii{logU} & -2\\
                                            & \ascii{zgas} & 0.02\\
                                            & \ascii{ne} & 100\\
                                            & \ascii{f\_esc} & 0.0, 0.02, 0.5\\
                                            & \ascii{f\_dust} & 0\\
                                            & \ascii{lines\_width} ($\rm km.s^{-1}$) & 300\\
		\hline
		\multirow{5}{16em}{\ascii{dustatt\_modified\_CF00}} & \ascii{Av\_ISM} & 0,0.1,0.2,0.5,1.0,1.5,2,2.5,3\\
                                                            & \ascii{mu} & 0.44\\
                                                            & \ascii{slope\_ISM} & -0.7\\
                                                            & \ascii{slope\_BC} & -1.3\\
                                                            & \ascii{filters} & V\_B90 \& FUV\\
		\hline
  		\multirow{4}{16em}{\ascii{dl2014}} & \ascii{qpah} & 2.5\\
                                            & \ascii{umin} & 1\\
                                            & \ascii{alpha} & 2\\
                                            & \ascii{gamma} & 0.1, 0.9\\
		
        \hline
  		\multirow{15}{16em}{\ascii{skirtor2016}} & \ascii{t} & 5, 7\\
                                                 & \ascii{pl} & 1\\
                                                 & \ascii{q} & 1\\
                                                 & \ascii{oa} & 10, 40, 80\\
                                                 & \ascii{R} & 20\\
                                                 & \ascii{Mcl} & 0.97\\
                                                 & \ascii{i} & 30, 70, 90\\
                                                 & \ascii{disk\_type} & 0 (Skirtor)\\
                                                 & \ascii{delta} & -0.36\\
                                                 & \ascii{fracAGN} & 0,0.1,0.2,0.5,0.8,0.99\\
                                                 & \ascii{lambda\_fracAGN} & 0/0\\
                                                 & \ascii{law} & 0 (SMC)\\
                                                 & \ascii{EBV} & 0,0.05,0.1,0.15,0.2,0.25,0.3,0.35,0.4,0.45,0.5,0.5,0.7,0.8,0.9,1\\
                                                 & \ascii{temperature} (K) & 100\\
                                                 & \ascii{emissivity} & 1.6\\
		\hline
  		\multirow{7}{16em}{\ascii{xray}} & \ascii{gam} & 1.9\\
                                        & \ascii{E\_cut} & 300\\
                                        & \ascii{alpha\_ox} & -1.9, -1.7, -1.5, -1.3, -1.1\\
                                        & \ascii{max\_dev\_alpha\_ox} & 0.2\\
                                        & \ascii{angle\_coef} & 0.5 \& 0\\
                                        & \ascii{det\_lmxb} & 0\\
                                        & \ascii{det\_hmxb} & 0\\
		\hline
	\end{tabular}
\end{table*}


\bsp	
\label{lastpage}
\end{document}